\documentclass[aps,letterpaper,twocolumn,preprintnumbers,floatfix,superscriptaddress]{revtex4}

\usepackage{graphicx}
\usepackage{dcolumn}
\usepackage{bm}
\usepackage{epsfig}
\usepackage{gensymb}
\usepackage{mathrsfs}  
\usepackage{amsmath}
\usepackage{amssymb}
\usepackage{graphicx,bm}
\usepackage{slashed} 
\usepackage{dsfont}
\usepackage[makeroom]{cancel}
\usepackage{youngtab}
\usepackage{multirow}
 \usepackage[titletoc]{appendix}

\def\fun#1#2{\lower3.6pt\vbox{\baselineskip0pt\lineskip.9pt
  \ialign{$\mathsurround=0pt#1\hfil##\hfil$\crcr#2\crcr\sim\crcr}}}

\def\lsim{\mathrel{\rlap{\raise 2.5pt \hbox{$<$}}\lower 2.5pt\hbox{$\sim$}}}
\def\gsim{\mathrel{\rlap{\raise 2.5pt \hbox{$>$}}\lower 2.5pt\hbox{$\sim$}}}

\input epsf

\newcommand{\be}{\begin{equation}}
\newcommand{\ee}{\end{equation}}
\newcommand{\bea}{\begin{eqnarray}}
\newcommand{\eea}{\end{eqnarray}}

\usepackage{color}

\newcommand{\comment}[1]{}

\begin{document}

\title{Trigonometric Parity for the Composite Higgs}

\author{Csaba Cs\'aki}
\affiliation{ Department of Physics, LEPP, Cornell University, Ithaca, NY 14853, USA}
\author{Teng Ma}
\affiliation{ 
 CAS Key Laboratory of Theoretical Physics, Institute of Theoretical Physics,
Chinese Academy of Sciences, Beijing 100190, China.}
\author{Jing Shu}
\affiliation{ 
CAS Key Laboratory of Theoretical Physics, Institute of Theoretical Physics,
Chinese Academy of Sciences, Beijing 100190, China.}
\affiliation{School of Physical Sciences, University of Chinese Academy of Sciences, Beijing 100049, P. R. China.}
\affiliation{CAS Center for Excellence in Particle Physics, Beijing 100049, China}
\begin{abstract}
We identify trigonometric parity as the key ingredient behind models of neutral naturalness for the Higgs potential. We show that any symmetric coset space readily includes such a trigonometric parity, which is simply a combination of a $\pi /2$ rotation along a broken direction and a Higgs parity transformation. We explain how to extend the top sector such that this $Z_2$ remains intact while the rest of the shift symmetry is explicitly broken, resulting in the cancelation of the quadratic divergences in the Higgs potential. Assuming additional structure (for example partial compositeness with maximal symmetry) can render the Higgs potential completely finite and with minimal tuning. We apply our principles to construct the minimal model realizing trigonometric parity based on $SO(6)/SO(5) \simeq SU(4)/Sp(4)$, yielding the simplest model of neutral naturalness. An added advantage of this model is that a simple fermionic UV completion can be easily identified. We analyze the tuning of the Higgs potential and find that the top partners can be quite heavy while vector mesons need to be relatively light to obtain minimal tuning. Finally we briefly comment on some novel phenomenology, including a possible six top final state at the LHC appearing in this model.  
\end{abstract}


\maketitle

\section{Introduction}
\label{Sec:intro}

What is the true nature of the Higgs boson? This question has bewildered theoretical physics ever since the Higgs boson was first proposed. One possible explanation first put forward in the 
80's~\cite{Kaplan:1983fs, Georgi:1984af, Dugan:1984hq}, and very extensively discussed over the past fifteen years~\cite{ArkaniHamed:2001ca,LH,Agashe:2004rs,SILH,Bellazzini:2014yua} is that the Higgs itself is not an elementary particle but rather composite, a strongly coupled bound state of some more fundamental elementary constituents. Such a scenario could explain why the Higgs was light if the Higgs (in addition to being composite)  was also a pseudo Nambu-Goldstone boson (pNGB) of a global symmetry spontaneously broken at energy scale $f$, described by the coset $G/H$. A composite Higgs remains one of the most fascinating options for electroweak symmetry breaking (EWSB) after the discovery of the 125 GeV Higgs.

The lightness of the Higgs boson also has a profound impact on the expected spectrum of beyond the Standard Model (BSM) particles: natural models of EWSB will predict the existence of light top partners. These can either be scalar top partners as in supersymmetric models, or fermionic top partners in composite Higgs models~\cite{LH,Contino:2006qr}. The top partners will cancel the bulk of the corrections to the Higgs potential (and allowing the Higgs to remain light).  The fermionic top partners in turn will produce the ``smoking gun" signals used for the LHC searches for colored top partners~\cite{DeSimone:2012fs, Li:2013xba, Burdman:2002ns,Perelstein:2003wd,Han:2005ru}. However, with the accumulated integrated luminosity already surpassing  60 fb$^{-1}$, our current bound on colored top partners is pushed up to 1 -- 1.2 TeV. The ever increasing top partner bounds may make one wonder whether there are 
options where the nice features of composite Higgs models are maintained without the existence of colored top partners.

Twin Higgs (TH) models present another interesting direction for stabilizing the Higgs potential~\cite{Chacko:2005pe,TH2, Craig}. In this scenario an additional $Z_2$ discrete symmetry is responsible for the cancellation of the quadratic divergences. In TH models the Higgs is also identified as a pNGB, and the $Z_2$ symmetry manifests itself via the $s_h \leftrightarrow c_h$ exchange symmetry in the Higgs potential~\cite{Chacko:2005pe, Geller:2014kta,Low:2015nqa, Barbieri:2015lqa}. This $Z_2$ is very efficient at softening the Higgs potential and eliminating most of the sources for tuning: in addition to canceling the quadratic divergences it also eliminates the so called ``double tuning" leading to Higgs potentials with minimal tuning. Furthermore this  $Z_2$ relates the top to the twin-top which is $SU(3)_c$ color neutral, thus also evading the bounds from direct top partner searches~\cite{ATLAS:2016btu, ATLAS:2016qlg, ATLAS:2016cuv}. While the TH framework is very attractive, the concrete models are not: the minimal $SO(8)/SO(7)$ coset space  is very large, leading to complicated models with very large representations. In these models the origin of the $Z_2$ exchange symmetry is not immediately obvious either.

In this paper, we introduce a new class of composite Higgs models with color neutral top partners,
where the origin of the $Z_2$ symmetry can be traced back to a simple and very generic discrete symmetry of the internal manifold describing the coset space. We argue that for any symmetric coset space the $s_h \leftrightarrow c_h$ exchange symmetry naturally emerges as a combination of Higgs parity with a $\pi /2$ rotation in the broken direction corresponding to the physical Higgs.  We will show how to extend this ``trigonometric $Z_2$ symmetry" such that it remains intact after the introduction of the top Yukawa couplings, which will provide a natural origin for the appearance of the color neutral top partners. The trigonometric $Z_2$ symmetry will relate the top and color neutral top partners to each other. For the gauge sector  we will not assume a twin mechanism:  indeed in all TH models the $Z_2$ twin parity is softly broken, and the breaking is usually assigned to the gauge sector. This will allow us to greatly simplify the group structure of the model, however additional ingredients will be needed to cancel the gauge contributions to the quadratic divergences (for example imposing the Weinberg sum rules, deconstruct a higher dimensional gauge theory, use the underlying UV strong dynamics or maximal symmetry). Using these principles we find the minimal model based the $SO(6)/SO(5)$~\cite{Gripaios:2009pe,Frigerio:2012uc} (or equivalently $SU(4)/Sp(4)$~\cite{Batra:2008jy, Galloway:2010bp,Cacciapaglia:2014uja}) coset, where the latter has a simple UV completion from fermion condensation. We also present several striking collider signatures, including novel six top final states.  

The structure of this paper is as follows. In Sec.~\ref{Sec:Z2} we discuss the origin of the NGB trigonometric parity, and then in Sec.~\ref{Sec:Realization_Z2} we show how to realize this trigonometric parity in the top sector of the simplest $SO(6)/SO(5)$ composite Higgs model (CHM) and also for general coset spaces and matter. In Sec.~\ref{Sec:SO_model}, we describe the deatails of the minimal $SO(6)/SO(5)$ CHM with trigonometric parity. In Sec.~\ref{Sec:UV_completion}, we present a UV completion through fermion condensation for the $SU(4)/Sp(4)$  (which is locally isomorphic to $SO(6)/SO(5)$). We also present the realization of trigonometric parity in this coset space, together with a possible UV completion for partial compositeness with $Z_2$ symmetry in this model. In 
Sec.~\ref{Sec:V_f} we calculate the Higgs potential from the fermion sector. In Sec.~\ref{Sec:tuning}, we discuss the structure of the entire Higgs potential and the tuning in this model and present the numerical results. In Sec.~\ref{Sec:pheno}, we briefly discuss some of the most strking phenomenological consequences of the minimal model. In Sec.~\ref{Sec:conclusion} we conclude and comment on the outlook. The Appendices contain some useful details on the $SO(6)$ and $SU(4)$ algebra, how to realize the pNGB trigonometric parity in gauge sector, the Higgs potential from the gauge sector, as well as how the UV completion guarantees that the Weinberg sum rule is satisfied leading to a finite Higgs potential.

\section{Origin of the trigonometric $Z_2$ symmetry}
\label{Sec:Z2}

Next we present our essential new observation:  a $Z_2$ symmetry useful for building TH-type  models is readily present for every Goldstone boson as long as a Higgs parity $V$ is maintained by the coset space. Such a Higgs parity automatically emerges for so-called symmetric coset spaces (which include most of the commonly used examples).  The reason for the appearance of such a $Z_2$ symmetry is quite simple: whenever we have a broken symmetry there is a  shift symmetry on the correspondig pion $\pi^i$ of the form $\pi^i/f \to \pi^i/f +\epsilon^i$. The effect of the Higgs parity is to simply reverse the sign of the pion $\pi^i \to -\pi^i$. Thus combining a $\pi/2$ rotation in the broken direction with Higgs parity will have the effect 
\begin{equation}
\label{eq:pichange}
\frac{\pi^i}{f} \to -\frac{\pi^i}{f}+\frac{\pi}{2}
\end{equation}
which on the trigonometric functions is equivalent to 
\bea \label{eq:pi_Z2}
  \sin  \frac{\pi^i}{f}  \leftrightarrow \cos \frac{\pi^i}{f} \ .
\eea
We call this the trigonometric $Z_2$ symmetry which is exactly the type of exchange symmetry one needs for the TH models~\cite{Chacko:2005pe} to cancel the quadratic divergences and also further reduce the tuning of the Higgs potential. It is automatically contained in every symmetric coset space, for example $SO(N+M)/SO(N)\times SO(M)$ and $SU(N+M)/SU(N) \times SU(M) \times U(1)$. Whether this symmetry will actually be realized on the Higgs potential will then depend on the structure of the explicit breaking terms. The task is to design the explicit breaking terms such that they break the general shift symmetry (in order to allow the generation of a Higgs potential) but maintain the $Z_2$ discrete subgroup of the shift symmetry identified above.  Once this is achieved the generated Higgs potential will be automatically exchange symmetric.  
 
To see this more explicitly we can consiser the coset space $SO(N+1)/SO(N)$ as a simple example.  This coset is equivalent to the rotations of an N-dimensional sphere $S^N$ because the nonlinear-pNGB field $U$ just describes the rotations of an $N+1$ dimensional unit vector. Hence the $N$ pNGBs can be identified with the rotation angles of $S^N$. The shift symmetry of any given $\pi^i$ can be thought of as the  $SO(2)_{i,N+1}$ rotation in a 2-dimensional plane in the $i^{th}$ and $N+1^{st}$ direction (if the vacuum is chosen in the $N+1^{st}$ direction $\mathcal{V} =(0,0,0,...,1)$) .  
In this case the $Z_2$ symmetry for pNGB $\pi^i$ is the $SO(2)_{i,N+1}$ rotation with ${\pi}/{2}$ angle combined with the Higgs parity transformation $V=\mbox{diag}(1,1,1....-1)$:    
 
\bea \label{eq:Z2_operator}
P_1 =e^{i\frac{\pi}{2} T^{\hat{i}} } V, 
\eea
where $T^{\hat{i}}$ is the broken generator associated with $\pi^{i}$. The linearly realized pNGB field can be expressed as $\Sigma = U\mathcal{V}$, where $U$ is the non-linear Goldstone field. Using the identity $P_1 \Sigma = e^{i\frac{\pi}{2} T^{\hat{i}} } V U V \mathcal{V} =e^{i\frac{\pi}{2} T^{\hat{i}} }U^\dagger \mathcal{V}$, we find that (as expected) $\pi^i$ transforms under $P_1$ as $\pi^i \to -\pi^i +\frac{\pi}{2}$, leading to the $\sin  \frac{\pi^i}{f}   \leftrightarrow \cos \frac{\pi^i}{f} $    $Z_2$ symmetry for $\pi^i$. For example for the case of $\pi^N$ the explicit expression of the Goldstone matrix $U$ 
\begin{equation}
U = \left( \begin{array}{ccc} \mathds{1}_{N-1} & & \\ & \cos \frac{\pi}{f} & \sin \frac{\pi}{f} \\
& -\sin \frac{\pi}{f} & \cos \frac{\pi}{f} \end{array} \right) 
\end{equation}
and the $P_1$ operator implementing the $Z_2$ on the Goldstone matrix is 
 \begin{equation}
P_1 = \left( \begin{array}{ccc} \mathds{1}_{N-1} & & \\  &  & 1 \\
& 1 &  \end{array} \right) 
\end{equation}
   
\section{Realizing the trigonometric $Z_2$ on the Lagrangian}
\label{Sec:Realization_Z2}

We have already seen that symmetric cosets always naturally contain the internal trigonometric $Z_2$ parity of Eq.~(\ref{eq:pi_Z2}). In order to make this useful we need to construct a Lagrangian that preserves this trigonometric $Z_2$ parity (while breaks the rest of the shift symmetry to allow the generation of a Higgs potential). As a simple and realistic ilustration we first present the top sector of the $SO(6)/SO(5)$ coset space. We will discuss the details of the gauge sector of the model later, for now all we need is that the $SO(4)$ containing the $SU(2)_L$ electroweak gauge group and $SU(2)_R$ custodial symmetry of the SM are embedded in the first four components of the $SO(6)$. In this case, the pNGB matrix $U$ corresponding to the physical Higgs boson will be given by  
\bea \label{eq:U}
U = \left(\begin{array}{cccc}
 \mathds{1}_3   &  &  &  \\ 
 & \cos \frac{h}{f}  &  & \sin \frac{h}{f} \\
 &  &  1 &  \\ 
 & -\sin \frac{h}{f} &  & \cos \frac{h}{f}
\end{array} \right). 
\eea
We can clearly see that the $4^{th}$ and $6^{th}$ rows and columns correspond to an $SO(2)$  rotation by angle $h/f$. As discussed above the shift symmetry for the Higgs is exactly this (broken) $SO(2)$ rotation. The explicit expression for the $Z_2$ trigonometric parity acting on the Higgs matrix (obtained by the combination of the $SO(2)$ rotation by angle $\pi /2$ with the Higgs parity transformation $V={\rm diag} (1,1,1,1,1,-1)$) is 
\bea
P_1^h=\left(\begin{array}{cccc}
 \mathds{1}_3   &  &  &  \\ 
 &   &  &1\\
 &  &  1 &  \\ 
 & 1&  &
\end{array} \right). 
\eea

Let us now consider the Yukawa couplings of the fermions. We embed the third generation SM quark  doublet $Q_L$ in the fundamental representation of $SO(6)$ while the right-handed top $t_R$ is assumed to be an $SO(6)$ singlet. The explicit expression for $Q_L$ using the standard embedding is
\bea \label{eq:embedding}
\Psi_{Q_L}&=&  \frac{1}{\sqrt{2}}  \left( \begin{array}{c}
b_L \\ 
-i b_L \\ 
t_L \\ 
i t_L \\ 
0\\
0
\end{array} \right) \ .
\eea                  
 
The top Yukawa coupling will be of the form 
\begin{equation}
\label{eq:SMYukawa}
y_t \bar{\Psi}_{Q_L} \Sigma t_R + h.c.
\end{equation}
Since the Higgs is composite there will be additional form factors showing up in Eq.~(\ref{eq:SMYukawa}) which however play no role in the following argument thus for simplicity we will suppress them for now. To extend the $Z_2$ trigonometric parity to the Yukawa couplings we must introduce the twin tops $\tilde{t}_{L,R}$ and an appropriate extension of the $Z_2$ parity involving the exchange of the ordinary and the twin tops. Due to the form of the embedding of $Q_L$ into $\Psi_{Q_L}$ we can see that the twin top also needs to be embedded into mutiple components on the SO(6) vector. This is the underlying reason why SO(6) is the smallest global symmetry where the trigonometric $Z_2$ can be implemented. Since version of the parity on the Higgs field involves exchanging the fourth and sixth components we embed the left handed twin top into the sixth component of an SO(6) vector. However, in the embedding of the ordinary top $t_L$ shows up twice, so the proper embedding of the twin top into an SO(6) vector will be. 
In order to realize the exchanging symmetry between $t$ and $\tilde{t}$ which contains the $P_1 ^h$ operation, the embedding for $\tilde{t}_{L}$ must be 
\bea
\label{eq:twintop}
\Psi_{\tilde{t}_{L}} =  \frac{1}{\sqrt{2}} \left( \begin{array}{c}
0 \\ 
0 \\ 
0 \\ 
0\\ 
\tilde{t}_L\\
 i\tilde{t}_L
\end{array} \right),
\eea  
while $\tilde{t}_R$ is also a singlet under $SO(6)$. We note that since we do not assume the existence of a twin $SU(2)_L$ gauge symmetry   $\tilde{t}_L$ and $\tilde{b}_L$ do not have to be in the same multiplet. We can now extend the Yukawa sector to include the twin top Yukawa coupling as well:
\begin{equation}
y_t \bar{\Psi}_{Q_L} \Sigma t_R + \tilde{y}_t \bar{\Psi}_{\tilde{t}_L} \Sigma \tilde{t}_R + h.c.\ .
\end{equation}
If $y_t = \tilde{y}_t$ this Lagrangian will be invariant under the trigonometric parity 
\begin{equation}
{\Psi}_{Q_L} \leftrightarrow P {\Psi}_{\tilde{t}_L}, \ \ t_R \leftrightarrow \tilde{t}_R, \ \ \Sigma \rightarrow P \Sigma
\end{equation} 
where $P$ is the parity operator implementing the exchange of  $t_L$ and $\tilde{t}_L$  
\bea
\label{eq:P}
P &=&P_0 P_1^h=  \left( \begin{array}{cccccc}
1&\\
&1&\\
& & &  &1 \\ 
& & && &1   \\ 
&&1\\
&&&1 \\
\end{array}  \right) \ .
\eea
In the above decomposition $P_0$ is the operator exchanging the $3^{rd}$  and $5^{th}$ 
components (and which acts trivially on the Higgs pNGB matrix) and also $[P_0, P_1^h]=0$.
Since this is a symmetry of the Lagrangian, the Higgs potential generated by these interactions must also be invariant under this trigonometric parity $P$. Since its action on the Higgs sector is 
$s_h \leftrightarrow c_h$, the Higgs potential must also be invariant under this exchange symmetry.

Now we can discuss the general construction for a trigonometric parity invariant top Yukawa sector. There always exists a $P_1^h$ trigonometric parity for the Higgs sector if the coset space $G/H$ is symmetric. We then introduce the twin tops $\tilde{t}_{L,R}$ and embed both the top and twin top into the same representation of $G$.  The exchange between top and twin top $t_{L,R} \leftrightarrow  \tilde{t} _{L,R}$ will induce a $Z_2$ operator $P$. The embedding of the twin tops should be chosen such that $P$ can be written as $P =P_0 P_1^h$ where $P_0$ is a trivial transformation on the Higgs matrix. If one then chooses the same form of the Yukawa couplings for the top and twin top sector the exchanges $t_{L,R} \leftrightarrow  \tilde{t} _{L,R}$ 
along with $s_h \leftrightarrow c_h$ will be a symmetry of the Yukawa sector, implying the $s_h \leftrightarrow c_h$ exchange symmetry for the generated Higgs potential. 

Is this exchange symmetry of the Higgs potential actually enough get rid of the quadratic divergences?
If the quadratic divergence is proportional to $s_h^2+c_h^2$ then it will be independent of the Higgs field and the quadratic divergences are eliminated. However in principle it could also be proportional to $s_h^4 + c_h^4$ which is still exchange symmetric  but would remain quadratically divergent. Which of these situations we encounter will depende on the representations chosen for the embedding for the top and twin tops. The actual Higgs potential depends both on the kinetic terms and the Yukawa couplings of the top sector. The quadratically divergent part actually only depends on either the Yukawa terms or the kinetic functions - any product of these would imply multiple insertions which soften the Higgs potential to be at most log divergent. Thus if the Yukawa contributions  and the kinetic contributions to the Higgs potential individually only depend on at most $s_h^2$ then the quadratic divergences will automatically cancel.  If the LH fermions are embedded into the fundamental representation of $G$ while the RH fermions into singlets then the Yukawa term will have only one power of $\Sigma$ insertion, while the kinetic terms at most two, and our condition will be satisfied.  We conclude that exchange symmetry in addition with chosing simple group representations will be sufficient for eliminating the quadratic divergences of the Higgs potential due to the top sector.

One interesting observation we want to mention here is that the $s_h \leftrightarrow c_h$ may actually also appear without introducing a mirror top $\tilde{t}$ but rather as a consequence of some symmetry purely within the top sector between left-handed top and right-handed top
\bea
\label{eq:Z2f2}
  t_{L,R}  \leftrightarrow {t} _{L,R} \; \mbox{or} \;   t_{L}  \leftrightarrow {t} _{R}  \quad s_h \leftrightarrow c_h. 
 \eea  
This is indeed the case for the maximally symmetric composite Higgs~\cite{Csaki:2017cep} or the left-right symmetric composite Higgs~\cite{LRCH}, both of which which render the Higgs potential finite with minimal universal tunning.

\section{The minimal $SO(6)/SO(5)$ model}
\label{Sec:SO_model}

We have seen in the previous section that the minimal coset which can implement the trigonometric $Z_2$ symmetry in the top sector (while incorporating an $SO(4)$ custodial symmetry) is $SO(6)/SO(5)$.  The purpose of this section is to provide the full description of the minimal model using the main ideas introduced previously.  One important point to emphasize is that the price of chosing the minimal coset $SO(6)/SO(5)$ suitable for implementing the $Z_2$ symmetry in the top sector is that the gauge sector will not be $Z_2$ symmetric - one can see that clearly from the embedding of the twin top into $\Psi_{\tilde{t}_L}$ in Eq.~(\ref{eq:twintop}). As a consequence the gauge contribution to the Higgs potential will be significant, and we will discuss that in more detail in App.~\ref{App:Gauge}. Besided being minimal another important advantage of the $SO(6)/SO(5)$ coset is that it is automorphic to $SU(4)/Sp(4)$ which has a simple UV completion, which is makes it one of the more desirable models of neutral naturalness. We will discuss the details of the UV completion in the next section. 

\subsection{The Goldstone/gauge Sector}

We start by explaining the structure of the gauge and Goldstone sector. The $SO(6)/SO(5)$ coset corresponds to $5$ NGBs parametrized by $h_{i}$ and $\eta$ with $i=1,2,3,4$.  The $SO(4)$ corresponding to  $SU(2)_L \times SU(2)_R$ of the SM electroweak group and custodial symmetry are contained in the $SO(5)$, and for simplicity we will choose the $SO(4)$ to correspond to the first four components of the $SO(6)$. The quatumn numbers of the Goldstones under the  
$ SU(2)_L \times SU(2)_R $ are~\cite{Gripaios:2009pe, Frigerio:2012uc}
\bea \label{eq:pNGBs}
 H \oplus \eta =  (2,2) \oplus (1,1) 
 \eea   
where $H$ will be identified with the SM Higgs doublet, $H=\frac{1}{\sqrt{2}}(h_2 + ih_1, h_4+i h_3)^T$. The non-linear Goldstone field is given by~\cite{Coleman:1969sm, Callan:1969sn} 
\bea
U=\mbox{exp} \left(i \frac{ \sqrt{2} \pi_{\hat{a}} T^{\hat{a} } }{f} \right),
\eea
where $\pi_{\hat{a}} =\{  h_i, \eta  \} $, $f$ is the decay constant, $T^{\hat{a}}$ are the broken generators corresponding to the NGBs. The generators are normalized as $\mbox{Tr}[T^{\hat{a}} T^{\hat{b}} ] =\delta^{{\hat{a}} {\hat{b}}} $, and are explicitly  listed in App.~\ref{App:Generator}. $U$ transforms non-linearly  under a global transformation ${\bf g} \in SO(6)$ as $U \to  {\bf g} U {\bf h}( \pi_{\hat{a}}, {\bf g})$  with ${\bf h} \in SO(5)$. As for the  $SO(5)/SO(4)$ minimal composite Higgs model (MCHM), we gauge $SU(2)_L$ and  $U(1)_Y \subset SU(2)_R$  to provide the electroweak gauge symmetries and in addition we also gauge the $SO(2)_\eta$ subgroup, corresponding to the rotations of the last two components of an $SO(6)$ vector. This $SO(2)_\eta$ is the broken direction providing the additional singlet Goldstone $\eta$. Since we gauge this direction, the $\eta$ will be eaten by the corresponding massive gauge boson.

The gauge interaction of the pNGB fields is most conveniently  written in terms of the $\Sigma$ field which is defined as $\Sigma = U {\cal V}$ where ${\cal V}=(0,0,0,0,0,f)$ is the VEV breaking $SO(6)$ to $SO(5)$. The $\Sigma$ transforms as $\Sigma \to {\bf g} \Sigma$ for any $g\in G$. The leading Goldstone Lagrangian is then given by 
\bea
\mathcal{L}=\frac{f^2}{2} (D_\mu \Sigma )^T D^\mu \Sigma,
\eea 
where $D_\mu = \partial_\mu  -i g W_\mu ^a T^a_L -i g^\prime B_\mu T_R ^3 -i g_1 B_\mu ^\prime T_\eta $ with $T_L ^a$, $T_R ^3$ and $T_\eta$ as generators of $SU(2)_L$, $U(1)_Y$ and $SO(2)_\eta$ embedded in $SO(6)$ (see App.~\ref{App:Generator}). After electroweak symmetry breaking, $\langle h \rangle \neq 0$,  the masses of SM and hidden gauge bosons are 
\bea \label{eq:boson_mass}
m_W ^2 =\frac{g^2 f^2}{4} s_h ^2, m_Z^2 =\frac{m_W ^2}{\cos ^2 \theta_W}, m_{B^\prime} ^2 = \frac{g^2 _1 f^2 (1- s_h ^2)}{2} ,
\eea 
where $\theta_W$ is the usual weak mixing angle. 
From Eq.~(\ref{eq:boson_mass}), the Higgs couplings to SM vectors $W^\pm_\mu / Z_\mu $ and $ B_\mu ^\prime$ can be extracted 
\bea
g_{h W^+_\mu W^-_\mu/ Z_\mu Z_\mu} &=& g_{hW^+_\mu W^-_\mu/Z_\mu Z_\mu} ^{SM} \sqrt{1-s_h^2}  \nonumber \\
 g_{h B_\mu ^\prime B_\mu ^\prime} &=& -\frac{2g_1 ^2  }{g^2} g_{hW^+_\mu W^-_\mu} ^{SM} \sqrt{1-s_h^2},    
\eea  
where $g_{h W^+_\mu W^-_\mu/ Z_\mu Z_\mu} ^{SM}$ is the Higgs coupling to $W^\pm_\mu / Z_\mu$ pairs in SM. 
   
\subsection{The Top and Bottom Sector}

 Next we discuss the fermion sector in detail. We have already presented the essential ingredients in Sec.~\ref{Sec:Realization_Z2}. Here we include all the form factors which will be present due to the strong dynamics leading to the global symmetry breaking and the composite Higgs, as well as  the construction of the bottom Yukawa couplings in a $Z_2$ invariant manner. 

The general low-energy effective Lagrangian of top-twin top-Higgs sector after integrating out the heavy fields is of the form~\cite{Csaki:2017cep}
\bea
\label{eq:genLag} 
\mathcal{L} &=& \bar{\Psi}_{Q_L} \slashed p (\Pi_0 ^q(p)  +\Pi_1 ^q(p) \Sigma \Sigma^\dagger) \Psi_{Q_L}+\bar{t}_R \slashed p \Pi_0 ^t(p) t_R \nonumber \\
&+& M_1^t(p) \bar{\Psi}_{Q_L}  \Sigma t_R \nonumber \\
&+& \bar{\Psi}_{\tilde{t}_L}  \slashed p (\tilde{\Pi}_0 ^q(p)  +\tilde{\Pi}_1 ^q(p) \Sigma \Sigma^\dagger) \Psi_{\tilde{t}_L}+\bar{\tilde{t}}_R \slashed p \tilde{\Pi}_0 ^t(p) \tilde{t}_R \nonumber \\
&+&\tilde{M}_1^t(p) \bar{\Psi}_{\tilde{t}_L} \Sigma \tilde{t}_R,    
\eea 
where   $\Pi_{0,1} ^{q}$($\tilde{\Pi}_{0,1} ^{q}$), $\Pi_0 ^t$($\tilde{\Pi}_0 ^t$)  and $M_{1}^t$($\tilde{M}_{1}^t$) are the form factors encoding the effect of the strong dynamics. We can see that there is an additional requirement for the $Z_2$ exchange symmetry: the form factors in the visible and twin sectors should be equal:
\bea 
\label{eq:identity}
\Pi_{0,1} ^{q} (p)=\tilde{\Pi}_{0,1} ^{q}(p), ~ \Pi_{0} ^{t} (p)=\tilde{\Pi}_{0} ^{t}(p),~ M_{1}^t (p)=\tilde{M}_{1}^t (p),
\eea 
which is expressing the requirement that the structure of the underlying strong dynamics should also be $Z_2$ symmetric. We will require in addition the condition that QCD and mirror QCD should be $Z_2$ symmetric
\bea
SU(3)_c \leftrightarrow  SU(3)_c ^\prime
\eea
otherwise QCD running effects will be different in the visible and the twin sectors, that could lead to significant (two-loop) corrections to the Higgs mass. 
 
Once the form factor relations Eq.~(\ref{eq:identity}) are satisfied, the effective Lagrangian has a  global $SO(6) \times SU(6)$ invariance where QCD and twin QCD are contained in the $SU(6)$:  $SU(3)_c \times SU(3)_c ^\prime \subset SU(6)$. So the $Z_2$ invariant effective Lagrangian can be written in the $SO(6) \times SU(6)$ invariant form
 \bea \label{eq:effective_SU6} 
\mathcal{L}_{eff} ^t&=& \bar{\Psi}_{L} \slashed p(\Pi_0 ^q(p) + \Pi_1 ^q(p) \Sigma^\dagger\Sigma) \Psi_{L} \nonumber \\
&+&   \bar{\Psi}_{R} \slashed p \Pi_0 ^t(p)  \Psi_{R} +   \bar{\Psi}_{L}  M_1 ^t(p) \Sigma \Psi_{R} +h.c, 
\eea
 where SM quarks and hidden fermions are embedded in ${\bf (6,6)}$ and ${\bf (1,6)}$ representation of  $\mathcal{G}' \equiv SO(6) \times SU(6)$ respectively, $\Psi_L =(\Psi_{Q_L}, \Psi_{\tilde{t}_{L}})$ and $\Psi_R =(t_R, \tilde{t}_{R} )$.

The effective Lagrangian can be written explicitly in terms of SM quarks  and hidden fermion $\tilde{t}$,
\bea 
\label{eq:top_effective}
\mathcal{L}_{eff} ^t&=& \bar{b}_L \slashed p \Pi_0 ^q(p) b_L +  \bar{t}_L \slashed p( \Pi_0 ^q(p) +\Pi_1 ^q(p) c_h ^2 )t_L   \nonumber \\ 
&+& \bar{t}_{R} \slashed p \Pi_0 ^t(p)  t_{R}+  \bar{\tilde{t} }_L \slashed p( \Pi_0 ^q(p) +\Pi_1 ^q(p) s_h ^2 )\tilde{t}_L   \nonumber \\
&+&  \bar{\tilde{t}}_{R} \slashed p \Pi_0 ^t(p) \tilde{t}_{R} - \frac{i M_1 ^t(p) }{\sqrt{2}} (\bar{t}_L t_R s_h +\bar{\tilde{t} }_L \tilde{t}_R c_h) +h.c. \nonumber \\
\eea  
We can see that this Lagrangian is $Z_2$ invariant.

For the bottom sector, we can introduce the left-handed twin bottom $\tilde{b}_L$ which is $SU(3)_c ^\prime$ triplet as
\bea
\Psi_{Q_L} ^\prime =\frac{1}{\sqrt{2}}  \left( \begin{array}{c}
t_L \\ 
-i t_L \\ 
b_L \\ 
i b_L \\ 
0\\
0
\end{array} \right) \quad \Psi_{\tilde{b}_{L}} =  \frac{1}{\sqrt{2}} \left( \begin{array}{c}
0 \\ 
0 \\ 
0 \\ 
0\\ 
\tilde{b}_L\\
 i\tilde{b}_L
\end{array} \right).
\eea    
such that the Higgs potential from the bottom sector has the $s_h \leftrightarrow c_h$exchange symmetry as a result of the $b \leftrightarrow \tilde{b}$ exchange.

The Higgs potential contributions from the fermion sector that break the $Z_2$ symmetry must be proportional to $|\epsilon_t|^2 |\epsilon_b| ^2$, where $\epsilon_{t,b}$ are the characteristic Yukawa couplings in the top and bottom sectors. Based on power counting, this potential is still log divergent. However the bottom Yukawa couplings are much smaller than top Yukawas, $\epsilon_b \ll \epsilon_t $, so the leading contributions to the Higgs potential will arise from the $Z_2$ preserving top sector $\sim \mathcal{O}(\epsilon_t^4)$ while the  $Z_2$ breaking  terms $\sim  \mathcal{O}( |\epsilon_t|^2 |\epsilon_b| ^2)$ which can be neglected. Thus in the following we only focus on the $Z_2$ invariant potential from top sector.

\section{$SU(4)/Sp(4)$ UV completion}
\label{Sec:UV_completion}

One of the main advantages of the minimal model presented above is that it has a simple UV completion. This is based on the fact that locally the cosets $SO(6)/SO(5)$ and $SU(4)/Sp(4)$ are isomorphic (see App.\ref{App:Map}) and the $SU(4)/Sp(4)$ coset can be realized via fermion condensation in a UV complete hypercolor theory. Here we focus on the underlying strong dynamics, particle content and the realization of the $Z_2$ symmetry in this coset.

\subsection{UV completion of the Higgs sector} 
In order to realize the $SU(4)/Sp(4)$ breaking pattern, we introduce four Weyl fermions $\psi_i$ with $i=1,2,3,4$~\cite{Galloway:2010bp,Cacciapaglia:2014uja}.  These preons will transform in the fundamental representation of the hypercolor gauge group $Sp(2N)$ (or alternatively could also be in the  spinor representation of a different hypercolor gauge group $SO(2N+1)$)~\cite{Ferretti:2013kya}. In this work we only focus the $Sp(2N)$ case. The electroweak gauge symmetries as well as the extra $U(1)_\eta \cong SO(2)_\eta$ are embedded in the global symmetry in the following way: the fermions $(\psi_1, \psi_2)$ are arranged into an $SU(2)_L$ doublet while the other two fermions, $\psi_3$ and $\psi_4$, are $SU(2)_L$ singlets. Their quantum number (including the two $U(1)$ charges) are summarized in Tab.~\ref{table:Quantum_Number}.~\footnote{This matter content has an $SU(2)_L^2 U(1)_\eta$ gauge anomaly which can be cancelled by introducing the hypercolor singlet $SU(2)_L$ doublet Weyl fermions $(\tilde{\psi}_1 , \tilde{\psi}_2 )$ with $U(1)_\eta$ charge $-2N$ and the $SU(2)_L$ singlet Weyl fermions $\tilde{\psi}_{3,4}$ with $U(1)_\eta$ charge $2N$ and hypercharge $\pm {1}/{2}$. }.
\begin{table}[htp]
\begin{center}
\begin{tabular}{c||ccccc}
&  $Sp(2N)$ & $SU(2)_L$ &  $U(1)_Y$&   $SU(3)_c$  &$U(1)_\eta$  \\
  \hline \\
$ ( \psi_1, \psi_2)$   & ${\tiny{\yng(1)}} $ &  ${\tiny{\yng(1)}}$  & ${\bf  0}$& ${\bf 1 }$ & $\bf 1$  \\ \\
 $\psi_3$ & ${\tiny{\yng(1)}}$ & ${\bf 1}$ &  ${\bf -\frac{1}{2} }$ & ${\bf 1 }$ & $\bf -1$  \\ 
&&& &\\
 $\psi_4$ & ${\tiny{\yng(1)}}$   &${\bf 1}$ &  ${\bf \frac{1}{2} }$& ${\bf 1 }$&$\bf -1$  \\  \\
 \hline 
\end{tabular}
\end{center}
\caption{ The quantum number of the Weyl fermion preons under gauge symmetries $ Sp(2N)\times SU(2)_L \times U(1)_Y \times SU(3)_c \times U(1)_\eta$}
\label{table:Quantum_Number}
\end{table}

Thus if the $Sp(2N)$ hypercolor group confines and the fermionic preons condense   $\langle \psi_i \psi_j  \rangle \neq 0$, similar to the QCD quark condensates, the $SU(4)$ global symmetry will be broken to its $Sp(4)$ subgroup, producing five NGBs.  If the vacuum has the form   
 \bea
 \label{eq:vacuum}
 V = \left( \begin{array}{cc}
i \sigma_2 & 0\\
0& -i \sigma_2 \\ 
\end{array}  \right).
 \eea   
the electroweak symmetries will be left unbroken by the preon condensates (while $U(1)_\eta$ will be one of the broken directions). 
  
By construction we are getting the exact same NGB pattern as in Eq.\ref{eq:pNGBs}. 
The preon content of these pNGBs are 
\bea
h_1&:& i( \psi_2 \psi_3+\psi_1 \psi_4-\psi_2^c \psi_3^c-\psi_1^c \psi_4^c )\nonumber \\
h_2&:&i(  \psi_2 \psi_3-\psi_1 \psi_4+ \psi_2^c \psi_3^c-\psi_1^c \psi_4^c )\nonumber \\ 
h_3&:& i( \psi_2 \psi_4-\psi_1 \psi_3- \psi_2^c \psi_4^c+\psi_1^c \psi_3^c) \nonumber \\
h_4&:&  \psi_1 \psi_3+\psi_2 \psi_4+ \psi_1^c \psi_3^c+\psi_2^c \psi_4^c \nonumber \\   
\eta &:& i( \psi_1 \psi_2+\psi_3 \psi_4- \psi_1^c \psi_2^c-\psi_3^c \psi_4^c),
\eea 
where $c$ stands for charge conjugation.  

The resulting structure of the NGB's will be identical to those of the $SO(6)/SO(5)$ case due to the local isomorphism between the two cosets (App.~\ref{App:Map}). For completeness we show the explicit form of the Goldstone matrices in the $SU(4)/Sp(4)$ coset below. 
The NGBs can be described by the Goldstone matrix~\cite{Coleman:1969sm, Callan:1969sn}  
\bea
U=\mbox{exp} \left(i \frac{\sqrt{2} \pi_{\hat{a}} T^{\hat{a} } }{f} \right),
\eea
 where the broken generators are again normalized as $\mbox{Tr}[T^{\hat{a}} T^{\hat{b}} ] =\frac{1}{2}\delta^{{\hat{a}} {\hat{b}}} $, and explicitly given in App.~\ref{App:SU4_Sp4_generator}.  
 
Just as in the $SO(6)/SO(5)$ case, only one pNGB remains (after EWSB) which will be identified with the Higgs. In unitary gauge the explicit form of the Goldstone matrix is  
\bea \label{eq:U}
U =\left( \begin{array}{ccc}
 c^\prime  \mathds{1}_2    & i \sigma^2 h s^\prime  \\ 
 i \sigma^2 h s^\prime   & c^\prime  \mathds{1}_2
\end{array} \right),
\eea
 where $c^\prime=\cos \frac{h  }{2 f} $ and $s^\prime=\sin \frac{h }{2 f}$. In this coset space, Higgs parity operation is the V combined matrix transpose. So, in the chosen vacuum, the automorphism map of symmetric space SU(4)/Sp(4) can be constructed as:
\bea
T^a \to -V T^{aT} V^T  \quad  T^{\hat{a}}  \to  -V T^{\hat{a} T} V^T,
\eea   
where $T^{a}$ is unbroken generator. Using this automorphism map we can construct the linearly realized Goldstone field $\Sigma'$ as 
\bea
\Sigma^\prime =U^2 V.
\eea  
It transforms linearly under a global transformation ${g } \in SU(4)$ as $\Sigma^\prime \to g \Sigma^\prime g^T$.  The gauge sector will again break the $Z_2$ symmetry explicitly and its contribution to Higgs potential is the same as the one in $SO(6)/SO(5)$ model, so we will not separately discuss it again here.

\subsection{The $Z_2$ symmetric top sector in $SU(4)/Sp(4)$}
\label{section:top_sector}
 
Since the UV completion is most easily formulated on the $SU(4)/Sp(4)$ coset, it is useful to translate the results on the $Z_2$ invariant top Yukawa couplings to this language. 
According to the correspondence between $SU(4)/Sp(4)$ and $SO(6)/SO(5)$ (see App.~\ref{App:Map}), the SM quark doublet $Q_L$ and the left-handed hidden fermion $\tilde{t}_L$ should be embedded in the ${\bf 6}$ anti-symmetric representation of the global $SU(4)$ while  the right-handed top $t_R$ and $\tilde{t}_R$  can be singlets. The explicit embeddings for the left-handed fermions are 
\bea \label{eq:SU4_embedding}
\Psi_{Q_L} &=& \frac{1}{\sqrt{2}} \left( \begin{array}{cc}
 {\bf 0} &  Q \\
-Q^T & \bf{ 0}  \\  
\end{array}  \right) \; \mbox{and}  \;  \Psi_{\tilde{t}_L}   =  \frac{1}{\sqrt{2}} \left( \begin{array}{cc}
 i \tilde{t}_L \sigma^2 &  0 \\
0 & \bf{ 0}  \\  
\end{array}  \right) \nonumber \\
\mbox{or}\;   \Psi_{\tilde{t}_L}  &=&  \frac{1}{\sqrt{2}} \left( \begin{array}{cc}
 \bf{ 0}  &  0 \\
0 & i \tilde{t}_L \sigma^2 \\  
\end{array}  \right),
\eea
where all of these are four by four antisymmetric matrices and $Q =(Q_L, 0)$. There are two different embeddings for $\tilde{t}_L$, which are physical equivalent, and we choose the first one to work with. For the above embeddings, the SM fermions are $U(1)_\eta$ neutral while the twin fermions do carry a $U(1)_\eta$ charge. Just as for the $SO(6)/SO(5)$ case, the $Z_2$ symmetry requires that the elementary-composite mixing terms be invariant under the enlarged global $ SU(4) \times SU(6)$ symmetry.
 We can first write down the effective Lagrangian for the elementary fermions coupled to the pNGB Higgs sector to explicitly show the $Z_2$ symmetry:
\bea \label{eq:effective_SU4} 
\mathcal{L}_{eff} &=&\Pi_0 ^q(p) \mbox{Tr}[  \bar{\Psi}_{L} \slashed p  \Psi_{L} ] +\Pi_1 ^q(p) \mbox{Tr}[  \bar{\Psi}_{L} \Sigma^\prime ] \slashed p \mbox{Tr}[ \Psi_{L} \Sigma^{\prime \dagger}]      \nonumber \\
&+&   \bar{\Psi}_{R} \slashed p \Pi_0 ^t(p)  \Psi_{R} +  M_1 ^t(p) \mbox{Tr}[ \bar{\Psi}_{L}   \Sigma^\prime ] \Psi_{R} +h.c, 
\eea
where  $\Psi_L =(\Psi_{Q_L} ,\Psi_{\tilde{t}_L} )$ and $\Psi_R =( t_R , \tilde{t}_R)$ are in the ${\bf (6,6)}$  and ${\bf (1,6)}$ representations of $SU(4) \times SU(6)$, and again $\Pi_{0,1} ^{q}$, $\Pi_0 ^t$ and $M_{1}^t$ are form factors. As expected the effective Lagrangian is invariant under the $Z_2$ transformation
\bea \label{eq:Z2_SU4} 
\Sigma &\to & P_1 \Sigma  P_1 =\Sigma (s_h \leftrightarrow c_h) \nonumber \\
\Psi_L &\to &  P_1  \Psi_L P_1  =  \Psi_L(t_L \leftrightarrow \tilde{t}_L, b_L \to -b_L), \nonumber \\
\Psi_R &\to &  \Psi_R(t_R   \leftrightarrow \tilde{t}_R),  
\eea  
where the operator $P_1$ 
\bea
P_1 &=& \left( \begin{array}{cccc}
1\\
&&1&\\
& 1& &  \\ 
 &  & &-1 
\end{array}  \right).  
\eea
is an element of $SU(4)$. So the $Z_2$ is a subgroup of $SU(4) \times SU(6)$. 
The effective Lagrangian can be written explicitly in terms of SM quarks and twin quarks  $\tilde{t}$ as
\bea \label{eq:top_effective2}
\mathcal{L}_{eff} ^t&=& \bar{b}_L \slashed p \Pi_0 ^q(p) b_L +  \bar{t}_L \slashed p( \Pi_0 ^q(p) -2\Pi_1 ^q(p) s_h^2 )t_L \nonumber \\
 &+&  \bar{t}_{R} \slashed p \Pi_0 ^t(p)  t_{R} +  \bar{\tilde{t}}_{R} \slashed p \Pi_0 ^t(p) \tilde{t}_{R}
  \nonumber \\ 
&+&  \bar{\tilde{t} }_L \slashed p( \Pi_0 ^q(p) -2\Pi_1 ^q(p) c_h ^2 )\tilde{t}_L    \nonumber \\
& -& \sqrt{2} M_1 ^t(p)  \left(\bar{t}_L t_R  s_h +\bar{\tilde{t} }_L \tilde{t}_R c_h    \right) +h.c.
\eea     
We can again see that this Lagrangian is $Z_2$ invariant and is equivalent to the effective Lagrangian in the $SO(6)/SO(5)$ case in Eq. (\ref{eq:top_effective}) up to a redefinition of the form factors.

\subsection{UV completion for partial compositeness with $Z_2$ symmetry}
Until now we have argued that the appropriate gauge $Z_2$ symmetric gauge-Goldstone structure can be nicely UV completed in the $SU(4)/Sp(4)$ coset, and also presented the necessary low-energy fermionic Lagrangian for this case. What remains to be shown is how this low-energy effective Yukawa term can actually be generated in the UV complete theory. The expectation is that it arises after integrating out a composite top partner fermion (and twin top partner), which are mixing with the elementary top (and mirror top)~\cite{Kaplan:1991dc,Agashe:2004rs}. For this to happen, some of the preons must be colored (otherwise they could not form a colored top partner bound state). These colored preons could either be fermions or scalars. While the theory with the scalars is somewhat simpler, it can not produce a composite fermion in the ${\bf 6}$ antisymmetric of SU(4), and also would reintroduce the hierarchy problem. Thus we focus on the case with fermionic colored preons. 

 In order to produce fermionic bound states, the Weyl fermion $\chi_{L,R}$ and its twin partners $\tilde{\chi}_{L,R}$, which are in anti-symmetric representation of $Sp(2N)$~\cite{Ferretti:2013kya}, must be introduced. The reason they are antisymmetric is so they can form bound states with two $\psi_i$ preons which end up to be color fundamentals. To maintain the $Z_2$ symmetry in the resulting Yukawa couplings the definition of the $Z_2$ symmetry must be extended to act on these preons:
\bea
 \chi_{L,R} \leftrightarrow  \tilde{\chi} _{L,R}. 
\eea
The quantum number of these fields under hypercolor and SM gauge symmetry $Sp(2N) \times SU(3)_c ^\prime \times SU(3)_c \times SU(2)_L \times U(1)_Y$ is 
summarized in Tab.~\ref{table:All_Quantum_Number}. 
\begin{table}[htp]
\begin{center}
\begin{tabular}{c||ccccc}
   & $Sp(2N)$ & $SU(2)_L$ & $U(1)_Y$& $SU(3)_c$ & $SU(3)_c ^\prime$ \\
     \hline \\ 
 $\chi_L$ & $\tiny{\yng(1,1)}$  & ${\bf 1}$& $\frac{2}{3}$ & ${\tiny{\yng(1)}}$ & ${\bf 1 }$ \\ 
 &&&& \\
 $\chi_R^c$ & $\tiny{\yng(1,1)}$  & ${\bf 1}$& $-\frac{2}{3}$ & ${\tiny{\bar{ \yng(1) }}}$ & ${\bf 1 }$ \\  
 &&&&\\ 
  $\tilde{\chi}_L$ & $\tiny{\yng(1,1)}$  & ${\bf 1}$& ${\bf1}$ &  ${\bf 1 }$ & ${\tiny{\yng(1)}}$  \\ 
 &&&& \\
 $\tilde{\chi}_R^c$ & $\tiny{\yng(1,1)}$  & ${\bf 1}$ & ${\bf 1}$ & ${\bf 1 }$&  ${\tiny{\bar{ \yng(1) }}}$  \\  \hline  
\end{tabular}
\end{center}
\caption{The quantum numbers of the colored preons.}
\label{table:All_Quantum_Number}
\end{table}

With respect to the hypercolor gauge symmetry, these preons have a global $ SU(12) \times U(1)$ symmetry with $SU(3)_c \times SU(3)_c ^\prime \subset SU(12)$. Once the hypercolor interactions condense, additional condensates might be formed. One plausible scenario would be for the colored preons to condense with each other $\langle \chi_i \chi_j \rangle \neq 0$ with $\chi$ =$\{ \chi_{L}, \chi_R ^c, \tilde{\chi}_{L}, \tilde{\chi}_{R} ^c \}$. Since the colored preons carry two hypercolor indices, these condensates would be symmetric and thus the $SO(12)$ global symmetry $SU(12)$ would break into $SO(12)$, resulting in $77$ additional pNGBs. We assume that these symmetric  condensates preserve $SU(3)_c \times SU(3)_c ^\prime$  and the quantum number of these pNGBs under $SU(3)_c \times SU(3)_c ^\prime $ are
\bea
\bf 77&=& \bf (8,1)+(6,1) +(\bar{6},1)+(1,8)+(1,6) +(1,\bar{6}) \nonumber \\
    &+&\bf (3,3)+(\bar{3},3) +(3,\bar{3})+(\bar{3},\bar{3})+(1,1).  
\eea
Since some of the pNGBs are colored, their masses are constrained to be heavier than at least $1$ TeV. However there are several sources for masses for these pNGB's~\cite{Cacciapaglia:2015eqa}:
\begin{itemize}
\item [ $\bullet$ ] gauge interactions: Since the $SU(3)_c \times SU(3)_c ^\prime$ gauge symmetries explicitly break the global symmetry, gauge bosons loops will contribute to the potential to these pNGBs. 
\item [ $\bullet$ ] preon mass terms for the $\chi$'s: gauge and $Z_2$ invariant mass terms for the preons can appear
\bea
 \mathcal{L} \supset m_{\chi} \chi ^T\left( \begin{array}{cccc}
& \mathds{1}_3 &  & \\
\mathds{1}_3 &&&\\
& & & \mathds{1}_3  \\ 
 &  & \mathds{1}_3 &
\end{array}  \right) \chi+h.c.
\eea 
These mass terms also explicitly break $SU(12)$ global symmetry and the pNGBs will get a mass similarly to the pion masses due to the quark masses in QCD. 
\item [ $\bullet$ ] the fermion Yukawas: The Yukawa couplings for the elementary fermions are not global $SU(12)$ invariant, so these fermions loops will also contribute to the pNGBs potential.    
\end{itemize}
In general the pNGB masses due to the above sources depend on the hypercolor scale
 $\Lambda_{\chi}$. The explicit breaking gauge and Yukawa terms will be loop suppressed while the preon mass will give a contribution similar to the ordinary pion masses $m_{pNGB}^2 \propto m_{\chi} \Lambda_{\chi}$. By choosing the preon mass sufficiently large (but sill $m_{\chi} \ll \Lambda_{\chi}$) one can ensure that this will be dominant positive contribution, moving the additional pNGB's above their experimental bounds. One might worry that increasing the preon mass also increases the top partner mass which would then increase the tuning in the Higgs potential sector for usual composite Higgs models. Here however the Higgs potential is stablized by the twin top $\tilde{t}$ so the heavy top partners do not result in significant tuning. Thus eventually the only significant tuning in this model is from generating the little hierarchy between the Higgs VEV and the symmetry breaking scale $f$.  

We close this section by a discussion of the classification of the fermionic composites of the UV completion of our model. Including all preons the full set of global symmetries for this model is $G=SU(4) \times SU(12) \times U(1)$. The wave function and quantum numbers under the global $SU(4) \times SU(12)$  and the unbroken subgroup $Sp(4) \times SO(12)$ for the fermionic bound states are shown in Tab.~\ref{table:Top_Partner_Quantum_Number}.  As expected in 
Sec.~\ref{section:top_sector}, the bound states $\chi (\psi \psi) $ or $\chi(\psi^c \psi^c)$ can play the role of top partners or twin top partners to produce the necessary $Z_2$ invariant Yukawa couplings.    

\begin{table}[htp]
\begin{center}
\begin{tabular}{c|c|c}
   & $SU(4) \times SU(12)$ & $Sp(4) \times SO(12)$  \\
     \hline \\ 
 $\chi (\psi \psi)$ & ${\bf (6,12)}$  & ${\bf (5, 12)}$, ${\bf (1,12) }$ \\ 
 & & \\
$\chi(\psi^c \psi^c)$ &${\bf (6,12)}$ &   ${\bf (5, 12)}$, ${\bf (1,12) }$ \\
& & \\
$\psi (\chi \psi)   $ &  ${\bf (10,12)}$ & ${\bf (10, 12)}$  \\
&& \\
$\psi (\chi^c \psi ^c)$  &${\bf (1, \overline{12})}$ & ${\bf (1, 12)}$\\
&& \\
$\psi (\chi^c \psi ^c)$  &${\bf (15, \overline{12})}$ & ${\bf (15, 12)}$

\end{tabular}
\end{center}
\caption{Quantum numbers of the composite fermions under the  global symmetry $SU(4) \times SU(12)$ and the unbroken subgroup $Sp(4) \times SO(12)$.  We use brackets "(..)" to denote the Lorentz index contraction of the fermions in the leftmost column.}
\label{table:Top_Partner_Quantum_Number}
\end{table}

\section{Higgs potential from fermion loops} 
\label{Sec:V_f}
In the previous sections, we have already shown the effective Lagrangian for the top and twin top coupled to the pNGB Higgses which preserves the trigonometric parity. In this section, we present the explicit realization by introducing both the top partners and the twin top partners dressed by the pNGB Higgses which form the composite operators. These composite operators couple to the top and twin top through the elementary-composite mixing interactions. The $SO(6) \times SU(6)$ invariant Lagrangian for the elementary and composite fermions based on the CCWZ formalism~\cite{Coleman:1969sm, Callan:1969sn}  can be constructed as:  
\bea \label{eq:top}
\mathcal{L} &=& f\bar{\Psi}_L U( \epsilon_{ 5L}  \Psi_{5R}+\epsilon_{ 1L}  \Psi_{1R} ) +f \epsilon_{R} \bar{\Psi}_R \Psi_{1L} \nonumber \\
&+& M_5 \bar{\Psi}_{5L} \Psi_{5R} +M_1 \bar{\Psi}_{1L} \Psi_{1R} +h.c, 
\eea  
where the composite partners of SM quarks and $\tilde{t}$, $\Psi_{Q,S}$ and $\tilde{\Psi}_{Q,S}$, which are five-plet and singlet of $SO(5)$,  are embedded in multiplets $\Psi_{5,1} = (\Psi_{Q,S} , \tilde{\Psi}_{Q,S})$ with quantum number ${\bf (5,6)}$ or ${\bf (1,6)}$ under $SO(5)\times SU(6)$. 
 The explicit formula for these fermion resonances in gauge basis is
\bea
\label{eq:top_partners}
\Psi_{Q} &=&
\left(
\begin{array}{c}
 i B -iX_{5/3} \\
 B+X_{5/3} \\
 T+X_{2/3}\\
 -T+X_{2/3}\\
 i T^\prime _+ - i T_{-}^\prime  \\  
  0 
\end{array}
\right)  \quad  \Psi_{S} =
\left(
\begin{array}{c}
0 \\
 0\\
 0\\
 0\\
 0  \\  
   T^\prime _+ +  T_{-}^\prime 
\end{array}
\right)  \nonumber \\ 
\tilde{\Psi}_{Q} &=&  \left(
\begin{array}{c} 
i \tilde{B}_{-1} -i \tilde{X}_{1} \\
\tilde{B}_{-1}+\tilde{X}_{1}  \\
 \tilde{T}_0+\tilde{X}_{0}\\
 -\tilde{T}_0+\tilde{X}_{0}\\
 i \tilde{T}^\prime _+ - i \tilde{T}_{-}^\prime  \\  
  0 
\end{array}
\right) \quad \tilde{\Psi}_{S} =
\left(
\begin{array}{c}
0 \\
 0\\
 0\\
 0\\
 0  \\  
   \tilde{T}^\prime _+ +  \tilde{T}_{-}^\prime 
\end{array}
\right),  
\eea
where the top partners  $(T,B)$ and $(X_{5/3}, X_{2/3})$ are electroweak doublet with hypercharge $1/6$ and $7/6$ and $T_{+,-} ^\prime$ are electroweak singlet with positive and negative $U(1)_\eta$ charge respectively. The twin top partners $( \tilde{T}_0, \tilde{B}_{-1})$ and $(\tilde{X}_{1}, \tilde{X}_{0})$ are electroweak doublet with hypercharge $-1/2$ and $1/2$ and $\tilde{T}_{+,-} ^\prime$ are also electroweak singlet with positive and negative $U(1)_\eta$ charge respectively.
It is easy to find that Eq.~(\ref{eq:top}) is invariant under the following transformation
\bea \label{eq:Z2}
U &\to & P U P_2 ^\dagger =U(s_h \Leftrightarrow c_h) \nonumber \\
\Psi_L &\to &  P \Psi_L =  \Psi_L(t_L \Leftrightarrow \tilde{t}_L) \nonumber \\
\Psi_5 &\to &  P_2 \Psi_5  \quad \Psi_1 \to P_2 \Psi_1 =\Psi_1 \nonumber \\
t_R & \Leftrightarrow & \tilde{t}_R,
\eea  
where 
\bea
 P_2 = \left( \begin{array}{cccccc}
1&0&0&0&0&0\\
0&1&0&0&0 &0\\
0& 0&0& 0 & 1&0 \\ 
0 & 0 & 0&-1 &0 &0  \\ 
0&0&1&0& 0 &0 \\
0&0&0&0& 0 &1 \\
\end{array}  \right).\nonumber \\ 
\eea
Since $P_2$ is the element of $O(5)$ and it only acts on the composite sector which is definitely invariant under $O(5)$,  the global $SO(6)\times SU(6)$ invariant Yukawa interactions are exactly $Z_2$ invariant, consistent with our previous discussion. As discussed above, this $Z_2$ can only keep the Higgs potential free of quadratic divergence, but the log divergence remains. In order to make the Higgs potential completely finite we can impose maximal symmetry in Eq.~(\ref{eq:top}). We find that the condition for Eq.~(\ref{eq:top}) being maximally symmetris is that the composites $\Psi_{1}$ and $\Psi_{5}$ fill out a full SO(6) fundamental representation and the elementary-composite mixing terms and the composite fermion mass terms are fully $SO(6)$ invariant:
\bea
 \epsilon_{ 5L} ^2 =\epsilon_{ 1L}^2  \quad M_1^2 =M_5 ^2.
\eea  
Notice that in our case, there is no ``twisted" mass for the composite fermions like in 
Ref.~\cite{Csaki:2017cep} and the relative sign between Yukawa couplings $\epsilon_{ 5L}$ and $\epsilon_{ 1L}$, and the one between the masses of composite fermons $M_1$ and $M_5$, are unphysical. More details on this slightly different realization of maximal symmetry compared to~\cite{Csaki:2017cep} will presented elsewhere~\cite{MSCTH}. In later calculations, we choose $\epsilon_{ 5L}  =\epsilon_{ 1L} =\epsilon_L$ and $M_1 =M_5=M$ to realize the maximal symmetry without losing generality. 

After integrating out the composite resonances and imposing maximal symmetry ($\Pi_1 ^q =0$), the explicit expressions for the form factors are:
\bea \label{eq:form_factor}
\Pi_0 ^q  &=&1-\frac{\epsilon_L ^2 f^2}{p^2 - M^2} \quad \Pi_0 ^t  =1-\frac{\epsilon^2 _R f^2}{p^2 - M^2} \nonumber  \\  M_1 ^t &=& \frac{\epsilon_L \epsilon_R f^2 M }{p^2 -M^2}.
\eea
The top mass and hidden fermion mass are 
\bea
m_t = \frac{\epsilon_L  \epsilon_R f^2 M}{\sqrt{2} M_T M_{T_1} } s_h  \quad m_{\tilde{t}} =m_{t} (s_h \to c_h),
\eea
where the top partner masses are $ M_T =\sqrt{M^2 +\epsilon_L ^2 f^2 } $ and $M_{T_1} = \sqrt{M^2 +\epsilon_R ^2 f^2 } $. 
So the Higgs potential from fermion loops is~\cite{Coleman:1973jx, Csaki:2017cep}
\bea
V_f   = -2N_c \int \frac{d^ 4 p_E }{(2\pi)^4} \mbox{log}\left(1 +   \frac{ M_t ^2}{2 p_E ^2 \Pi_0^t  \Pi_0 ^q } s_h ^2\right) +(s_h \to c_h)   
\eea    
Since the Higgs potential from the top loops is invariant under $s_h \leftrightarrow c_h$, its leading order must be proportional to the fourth power of the top Yukawa  $|M_t|^4$. We can expand the Higgs potential in the  top Yukawa and obtain
\bea \label{eq:Vf}
V_f = \frac{2N_c}{(4\pi)^2} \int dp_E ^2 p_E ^2 \left( \frac{ M_t ^2}{4 p_E ^2  \Pi_0^t  \Pi_0 ^q } \right)^2( -s_h ^2  +s_h ^4 ). 
\eea    
It is interesting to find that the Higgs potential in this model is exactly the same as for the  Twin Higgs model based on $SO(8)/SO(7)$~\cite{Low:2015nqa, Barbieri:2015lqa, Geller:2014kta}.  The reason is that  the $Z_2$ symmetry realizing neutral naturalness is purely a subgroup of $SO(2)$. So the minimal composite Higgs model with custodial symmetry that realizes neutral naturalness is $SO(6)/SO(5)$. 

\section{Tuning in the Higgs potential and Numerical Scan}
\label{Sec:tuning}
\begin{figure}
\begin{center}
\includegraphics[width=0.8\columnwidth]{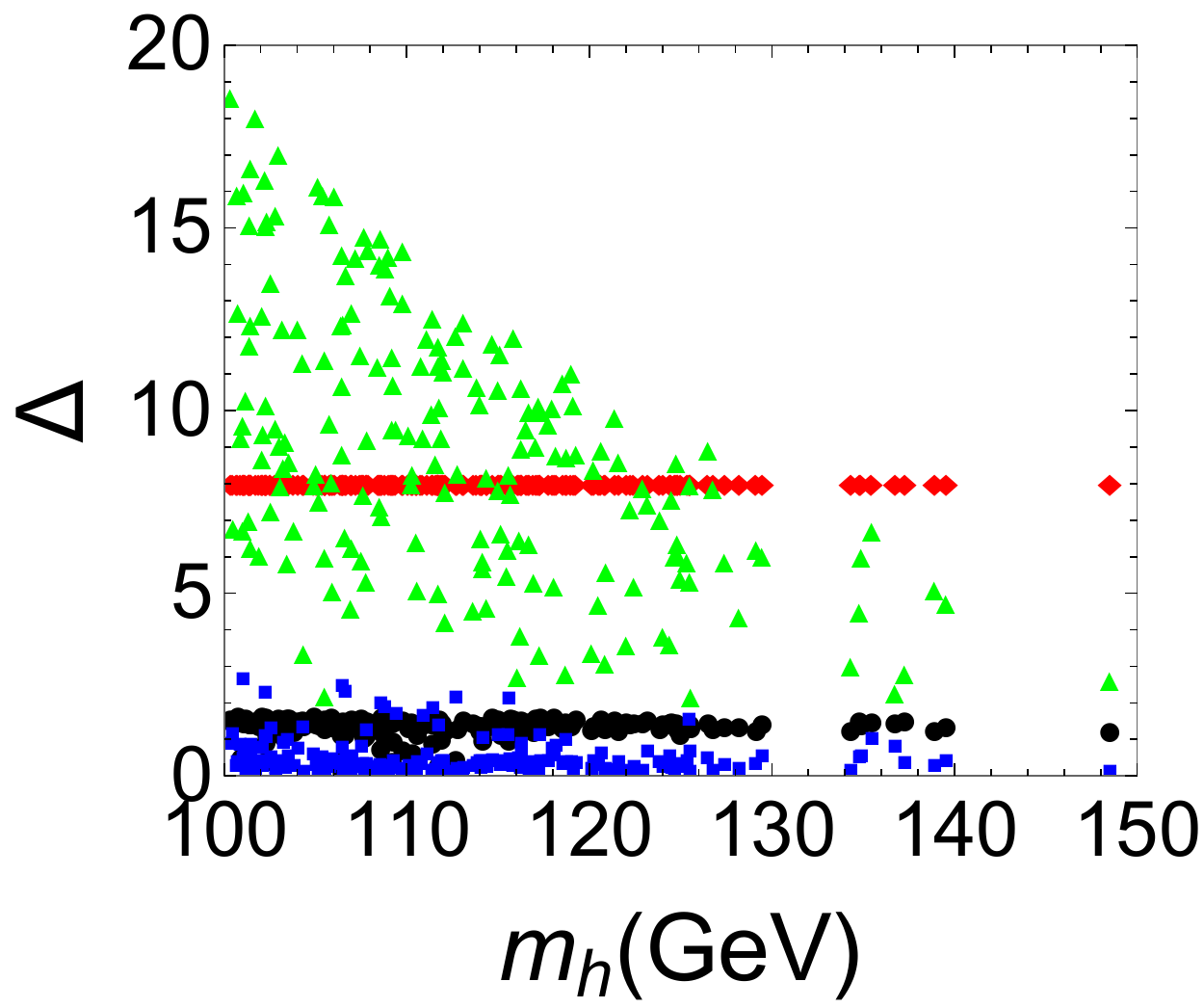}
\end{center}
\caption{Scatter plot of tuning $\Delta_i$ for the various input parameters $x_i$, $M$ (black), $\epsilon_{L}$ (blue), $\epsilon_R$ (red), $m_\rho$ (green) and $g_1$ (magenta), as a function of $m_h$ with $\xi =0.1$ held fixed. We choose the range of parameters as follows: $m_t \in [140, 170]$ GeV, 2$m_{B^\prime} >125$ GeV, $m_\rho \in [2,  3]$ TeV and $M >1$ TeV. } 
\label{fig:alltuning}  
\end{figure}

\begin{figure}
\begin{center}
\includegraphics[width=0.49\columnwidth]{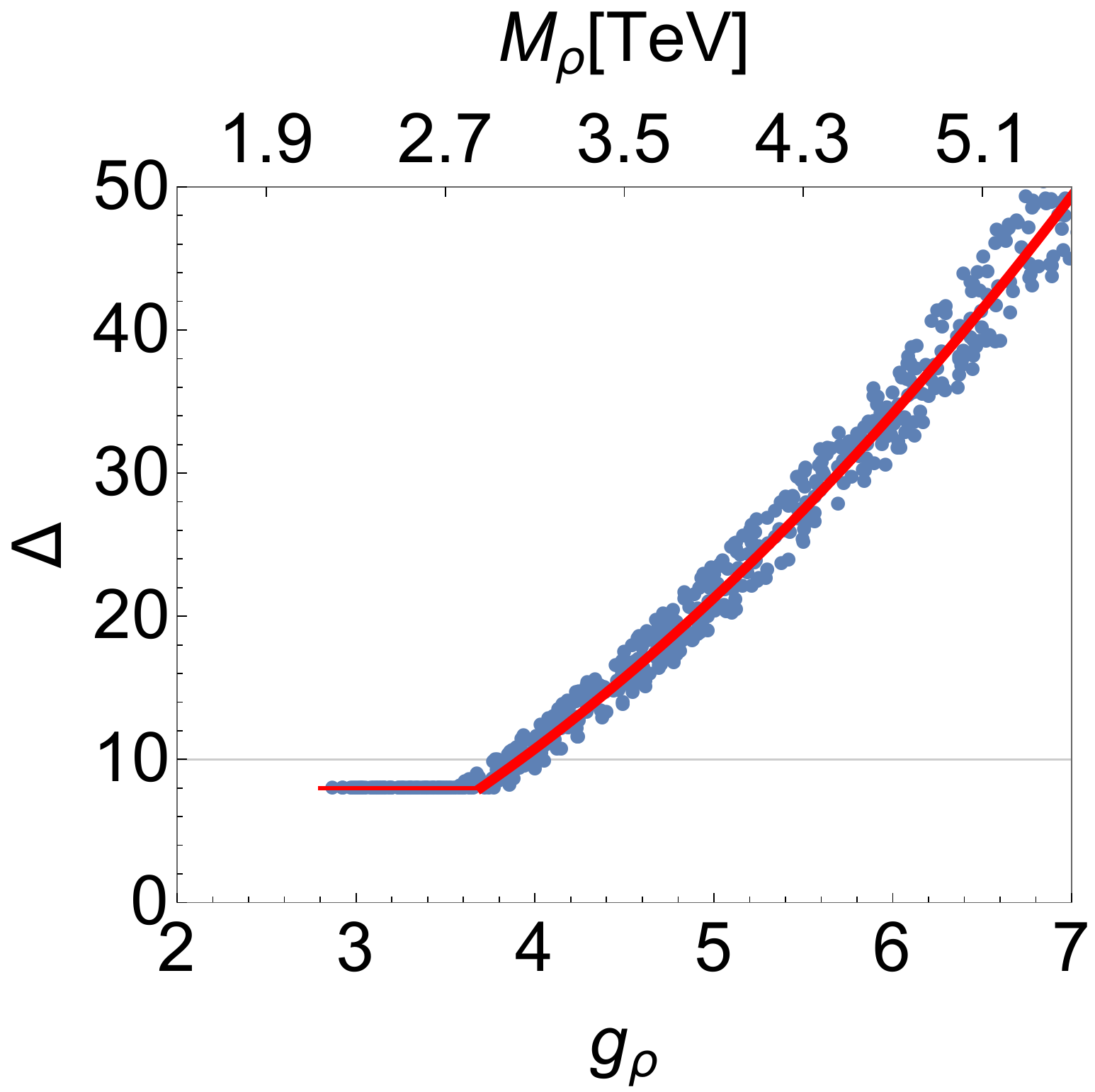}
\includegraphics[width=0.49\columnwidth]{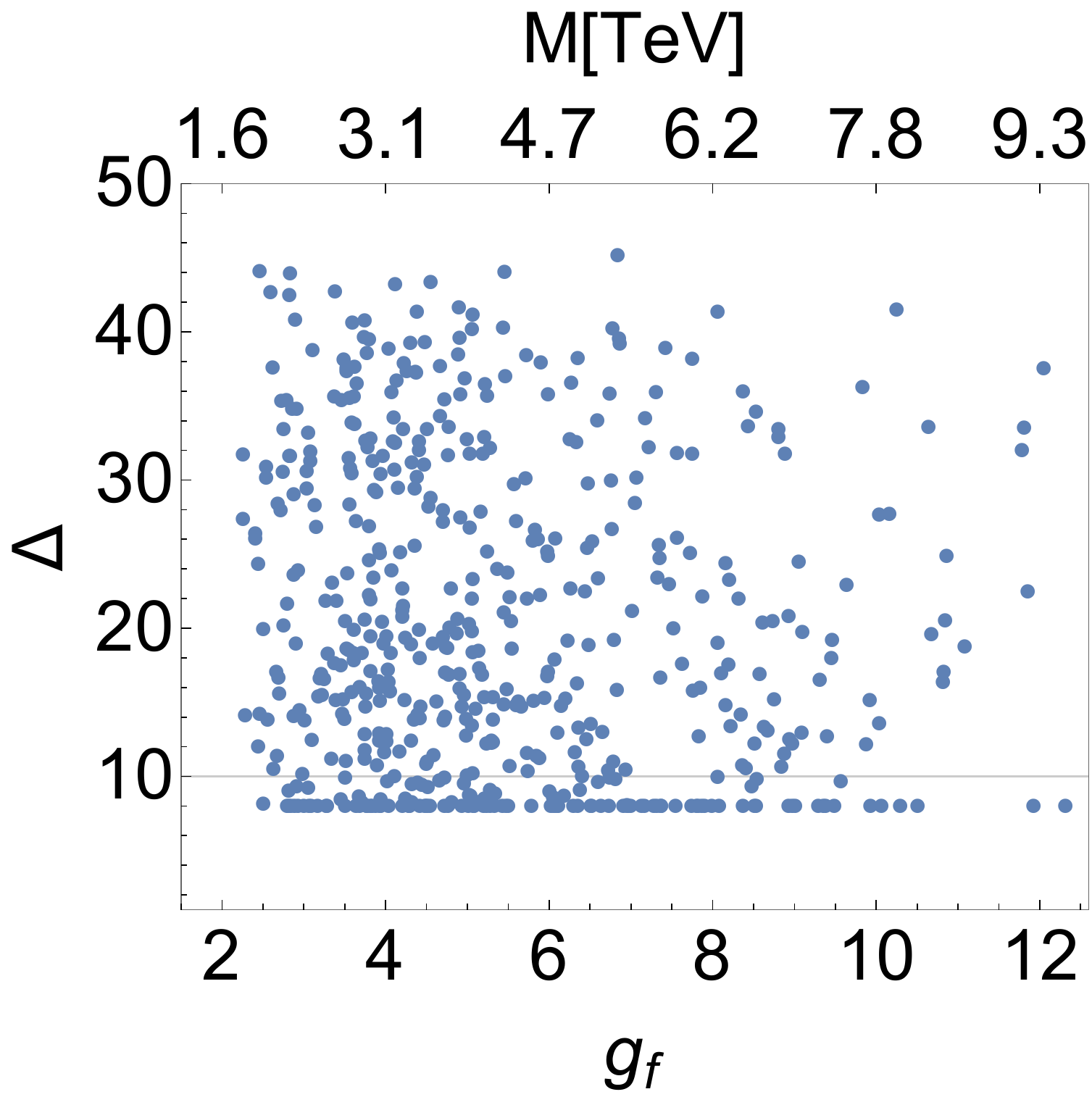}
\end{center}
\caption{Scatter plot of tuning $\Delta$  as a function of $g_\rho = m_\rho/f$(Left) and  vectorial mass $g_f = M/f$(Right) for $m_h =125$ GeV with $\xi =0.1$ held fixed. The chosen range of parameters is $m_t \in [140, 170]$ GeV, 2$m_{B^\prime} >125$ GeV, $m_\rho > 2$ TeV and $M > 1$ TeV. The gridline line corresponds to the minimal tuning $1/\xi $. } 
\label{fig:tuning}  
\end{figure}

The leading expression of the Higgs potential from gauge and fermion loops can be parametrized as 
\bea
V_g =\gamma_g s_h^2 \quad V_f =\gamma_f(-s_h^2 +s_h^4), 
\eea
where contributions from the gauge sector follow from Eq. (\ref{eq:Vg}) in App~\ref{App:Gauge} by assuming it is UV finite:
\bea 
V_g \simeq \frac{3f^2(3g^2+g^{\prime 2} -  2g_1 ^2 )m_\rho ^2 \ln 2}{64\pi^2} s_h ^2.
\eea
The total Higgs potential without considering higher dimensional operators in the UV is given by 
\bea
V=V_g +V_f =- \gamma s_h^2 +\beta s_h ^4,   
\eea
where $\gamma = \gamma_f -\gamma_g$ and $\beta =\gamma_f$(the contribution to $\beta$  from gauge bosons is at $\mathcal{O}(g^4)$ or $\mathcal{O}(g^{\prime 4})$ so it can be neglected.). 
The potential has a local minimum for $\gamma >0$
\bea
s_h ^2 =\xi =\frac{\gamma}{2\beta}.
\eea   
Similar to~\cite{Csaki:2017cep}, the potential from the fermion sector has a vacuum at $\xi =0.5$. In order to reduce $\xi$ to experimentally allowed values $\xi \ll 1$, the contribution from the gauge sector must be included and a cancellation between gauge and fermionic contributions in the $s_h ^2$ term  $\gamma_f \approx \gamma_g$ must be imposed. As in~\cite{Csaki:2017cep}, the tuning in this model  will be around the minimal tuning $\Delta \simeq 1/\xi$. In this model the Higgs potential $V_f$  is quartic in the top Yukawa coupling,  $\gamma_f \sim  \mathcal{O}(y_t^4)$,  so, according to the power counting, it is not explicitly dependent on the top partner masses and its explicit expression at leading order is~\cite{Panico:2012uw}   
 \bea
 V_f &\simeq &  c^\prime  \frac{N_c M_f ^4 }{16 \pi^2 }(\frac{y_t}{g_f} )^4[-s_h ^2 +s_h^4] \nonumber \\ 
       &\simeq & c^\prime \frac{N_c f ^4 }{16 \pi^2 } y_t^4[-s_h ^2 +s_h^4],
 \eea 
 where $c^\prime$ is an order one dimensionless constant,  $M_f$ is a typical fermion resonance mass and the associated coupling $g_f$ is defined by $g_f =M_f /f$. Thus the Higgs mass does not linearly depend on top partner mass and heavy top partners can be achieved without increasing the tuning. Since the Higgs potential is suppressed at $\mathcal{O}(y_t^4)$, a light Higgs can be easily produced without much tuning. To see this, we can explicitly estimate its mass: 
\bea
m_h^2 &=& \frac{8\beta}{f^2} \xi (1-\xi)   \nonumber \\
 &\Rightarrow &  \frac{m_h^2}{m_t^2} \simeq  c^\prime \frac{ y_t^2 N_c }{2\pi^2} (1-\xi) \approx 0.07(1-\xi)c^\prime.
\eea
So $m_h =125$ GeV can be naturally produced for a natural value $c^\prime \approx 8$ and the corresponding tuning is $1/c^\prime \ll 1$. 
To explicitly confirm our estimation, we numerically evaluate the tuning of this model. In this work we use the measure of tuning~\cite{Barbieri:1987fn} 
\bea
\Delta=\mbox{max} \{\Delta_i \}  \quad   \Delta_i  =  \frac{ \partial \mbox{log} \xi } {\partial \mbox{log} x_i},
\eea 
where $x_i$ is the free parameter of the theory.  In Fig.~\ref{fig:alltuning}, we show the tuning $\Delta_i$ for all free parameters as a function of $m_h$ for $\xi =0.1$ and a light vector meson $m_\rho \in [2,  3]$ TeV. The range of the parameters is taken as follows: $m_t \in [140, 170]$ GeV, 2$m_{B^\prime} >125$ GeV and $M >1$ TeV. One can see that the largest tuning is from the  cancellation between contributions from the gauge and top sectors. The tuning is large for light Higgs but remains constant for heavy Higgs. The analytical expression for the tuning $\Delta$ is 
\bea
\label{eq:tuning}
\Delta &\simeq &  \frac{2 \gamma_g}{|\gamma_f -\gamma_g|} =1 /\xi-2 \; \mbox{for}   \;  m_\rho <\frac{4\pi m_h \sqrt{1-2\xi} }{3g \sqrt{\mbox{ln} 2 \xi (1-\xi)} }  . \nonumber \\ 
\Delta &\simeq & \frac{1}{\xi}(1-2\xi) \left(\frac{9\mbox{ln}2\xi(1-\xi) g^2m_\rho ^2  }{8\pi^2 m_h^2(1-2\xi) } -1\right) \; \mbox{for} \;  \nonumber \\
m_\rho &>& \frac{4\pi m_h \sqrt{1-2\xi} }{3g \sqrt{\mbox{ln} 2 \xi (1-\xi)} } .  
\eea 
In Fig.~\ref{fig:tuning}, we show the tuning $\Delta$ as a function of $g_\rho ={m_\rho}/{f}$(Left) and vector fermion mass $g_f = { M}/{f}$(Right) for Higgs mass $m_h =125$ GeV with $\xi =0.1$. The red line is the analytical expression for the  tuning in Eq.~(\ref{eq:tuning}). Similar to the tuning in the composite Twin Higgs model with $Z_2$-breaking in the $SU(2)$ gauge sector~\cite{Low:2015nqa}, the left panel shows the minimal tuning is achieved for $g_\rho \lesssim 3.8$ and the tuning increases for $g_\rho > 3.8$ because the cancellation between the contribution from electroweak gauge bosons and hidden gauge bosons becomes significant. The right panel shows that both low tuning $\Delta \sim 8$ and heavy top partners can be achieved for a light Higgs. In the case of $Z_2$ breaking in $SU(2)$ gauge sector where the Higgs quadratic divergence from gauge bosons is killed by extra symmetries (like deconstruction in Ref.~\cite{Low:2015nqa}), the $SO(8)$ global symmetry is redundant.

Finally it is interesting to mention that for the case of $SU(4)/Sp(4)$ with UV completion, the underlying strong dynamics automatically enforces the 1st and 2nd Weinberg sum rule for vector resonances from strong dynamics (for a detailed discussion see App.~\ref{App:Gauge}). Therefore in this case, one does not need extra structure for the model to keep gauge contributions to the Higgs potential finite and leave the model simple and elegant~\footnote{The approximate $m_\rho \approx 4\pi f/\sqrt{2N}$ indicate that in order to obtain a small tuning from a light $\rho$, a relatively large $N$ is preferred.  }.

\section{some comments on Phenomenology}
\label{Sec:pheno}
In the minimal model which realizes neutral naturalness, some of the phenomenology is the same as for the $SO(8)/SO(7)$ composite twin Higgs model. In this section we briefly review the experimental constraints and demonstrate that our model is consistent with the current experimental bounds with the minimal tuning. We also discuss some unique phenomenology in this model. The full study of the phenomenology will be presented elsewhere.
\begin{itemize}
\item[$\bullet$] Signals related to the hidden sector: 
The Higgs can decay to hidden particles such as hidden gauge bosons $B^\prime_\mu$ and twin bottom $\tilde{b}$. The $B^\prime_\mu$, on the other hand, can decay into SM fermions through their composite components which are charged under $SO(2)_\eta$ ($U(1)_\eta$). Therefore, the coupling between $B^\prime_\mu$ and SM fermions are proportional to the size of elementary-composite mixings so $B^\prime_\mu$ will  decay into bottom for $m_{B^\prime_\mu} \lesssim 2m_t$ and to top for $m_{B^\prime_\mu} > 2m_t$. Therefore, If $m_{B^\prime_\mu} < m_h/2$, it can contribute to the Higgs decay width through the four bottom channel. In the bottom sector, the invisible decay of Higgs is dominated by decay to a pair of mirror bottom and it is order of $5-10 \% $ for $\xi =0.1$~\cite{Craig:2015pha}. 

\item[$\bullet$] Conventional direct searches: The top partner mass reach is $M \geq 2$ $(2.8)$ TeV at 14 TeV LHC with the integrated luminosity $300$ fb$^{-1}$ ($3$ ab$^{-1}$)~\cite{Matsedonskyi:2014mna}, which can be easily satisfied in our model with the minimal tuning. The $\rho$ meson mass reach is around $m_\rho \geq 3.6$ TeV at 14 TeV LHC with integrated luminosity $300$ fb$^{-1}$~\cite{Contino:2013gna}, therefore we not only can cover the minimal tuning parameter space but also some space with moderate tuning. The mirror top lies in the TeV range with mass $\sim y_t f$. Its mass reach is $\sim 200$ GeV at LHC with the integrated luminosity $3$ ab$^{-1}$ for pair production through off-shell Higgs~\cite{Craig:2014lda}, which does not impose any bounds in our model.  


\item[$\bullet$] Six tops signal: Some of the top partners $T^\prime _{+,-}$, which are electroweak singlet and charged under the $SO(2)_\eta$ ($U(1)_\eta$) gauge symmetry, mix with our SM top after electroweak symmetry breaking. By rotating into the physical mass basis, this kind of resonances denoted by $t^\prime _{+,-}$ can decay into three tops through the $SO(2)_\eta$ ($U(1)_\eta$) gauge interactions,
\bea
t^\prime \to B_\mu ^\prime \; t \to t \; \bar{t} \; t. 
\eea
So when they are pair produced at the LHC, a striking signal corresponding to six top final states is predicted. The background for the six top signal is very tiny, so this channel can impose significant bounds on heavy colored top partners. At present, there are no LHC searches for six tops and projections  from other searches like black hole~\cite{TheATLAScollaboration:2015dhj,Aad:2015mzg}, multi-lepton SUSY~\cite{Aad:2012ms}, etc are loose. A well designed strategy to search for six or multiple tops produced in LHC will be presented elsewhere~\cite{Sixtop}. Nevertheless, from the right panel of Figure~\ref{fig:tuning}, the mass of the colored top partners can be heavier than 10 TeV so this signal is not mandatory. 

\item[$\bullet$] Wess-Zumino-Witten (WZW) terms: If the $SO(2)_\eta$ ($U(1)_\eta$) is not gauged, the singlet $\eta$ can decay to $W_\mu$ or $Z_\mu$ gauge boson pairs through the WZW anomaly terms $\sim {\eta} W_{\mu \nu} ^a W_{\mu \nu}^a/f$ if the $U(1)_\eta$ global anomaly does not vanish, where $W_{\mu \nu}^a$ is the $SU(2)_L$ gauge field strength. For the $SU(4)/Sp(4)$ case with partial compositeness based on the fermionic UV completion, the actual global symmetry is $U(4) \times U(12)$. However the two $U(1)$ symmetries corresponding to the numbers of the two types of preons $\psi_i$ and $\chi_j$,are broken into a single global $U(1)$ symmetry by the hypercolor gauge interactions. Its mass is determined by the colored preon mass $m_{\chi}$ and the radiative corrections. This pNGB has  WZW terms with QCD gluons and electroweak gauge bosons, so it can have large production cross section through gluon fusion and decay into a gauge boson pair at the LHC. The detailed study of its phenomenology at the  LHC can be found in~\cite{Belyaev:2016ftv}.                                 
\end{itemize}
 
\section{Conclusion and outlook}  
\label{Sec:conclusion}

We have presented a novel approach to the $Z_2$ parity necessary to construct TH-type models. The key observation is that for arbitrary symmetric $G/H$ coset spaces a $Z_2$ symmetry naturally emerges which can remain unbroken after introducing the matter fields. We call this $Z_2$ the ``trigonometric parity", which is just a combination of the Higgs parity with a $\pi /2$ rotation in the broken direction.  Once the twin partners are introduced in a manner that preserves the trigonometric parity, the Higgs potential will have the $s_h \leftrightarrow c_h$ symmetry which renders the Higgs potential free of quadratic divergences. 

We construct concrete models based on the minimal coset $SO(6)/SO(5)$ and $SU(4)/Sp(4)$ (which is locally isomorphic to $SO(6)/SO(5)$. The advantage of the latter case is that it is easy to find a  UV completion for the Higgs sector and the top-twin top sector. For the top and twin top sector, we can in addition require maximal symmetry and partial compositeness to further soften the Higgs potential and make it finite. These tools can also be useful to arrive at the correct light Higgs mass. For a realistic model one needs additional spin-1 resonances to eliminate the quadratic divergences in the Higgs potential from the $W/Z$ sector  and are essential to provide the correct EWSB. These spin 1 partners may arise from additional symmetry structures or simply from the underlying strong dynamics. The twin tops are color singlets, therefore evade the strong constraints from direct LHC searches. The colored top partners are heavy and can even be outside of the LHC energy range. The SM and twin sectors communicates through the pNBG Higgs with $v/f$ suppression or the elementary-composite mixings. There can be interesting phenomenological cosnequences of the model presented here. One striking signal at the LHC could be final states with six tops from the cascade decay of the top partners.

Our results can be extended into several interesting directions.  
  For example we can turn on the electroweak preon mass $m_\psi$ to trigger the EWSB. The minimal $SO(6)/SO(5) \simeq SU(4)/Sp(4)$  structure can be realized in warped space. The additional $\psi_i$ preons of the $SU(4)/Sp(4)$ model can be confined at the UV brane.
We may also consider four fermion interactions between the tops and preons~\cite{Galloway:2010bp,Cacciapaglia:2014uja} instead of the partial compositeness in our setup. Finally, as already briefly mentioned before, trigonometric parity may be realized in the top sector without introducing twin tops, leading to either the maximal symmetry~\cite{Csaki:2017cep} or left-right symmetry~\cite{LRCH} to make the Higgs potential finite. These exciting connections can open a vast new field of unexplored model building  which we expect will lead to many more fruitful results in the future.  

\section*{Acknowledgements} 
C.C. and J.S. thank the Mainz Institute for Theoretical Physics for its hospitality while this work was in progress. C.C. also thanks the Aspen Center for Physics - supported in part by NSF PHY-1607611 - for its hospitality while working on this project. C.C. is supported in part by the NSF grant PHY-1719877. J.S. is supported by the NSFC under grant No.11647601, No.11690022 and No.11675243 and also supported by the Strategic Priority Research Program of the Chinese Academy of Sciences under grant No.XDB21010200 and No.XDB23030100. T.M. is supported in part by project Y6Y2581B11 supported by 2016 National Postdoctoral Program for Innovative Talents.   
  
\section*{Note added}
While this paper was being finalized, Ref.~\cite{Serra:2017poj} appeared which presents a similar  $SO(6)/SO(5)$ model.  
  
\appendix
\section{$SO(6)$ generators} 
\label{App:Generator} 
In this appendix, we list the generators of $SO(6)$. They can be classified in five broken ones in $SO(6)/SO(5)$, six unbroken generators of custodial symmetry $SO(4) \cong SU(2)_L \times SU(2)_R \in SO(5) $ and four unbroken ones of $SO(5)/SO(4)$ coset    
 \bea
 T^{ \hat{a} }_{ij} &=&  -\frac{i}{\sqrt{2}}(\delta^{\hat{a} i} \delta^{6j} -\delta^{\hat{a} j} \delta^{6i}), \nonumber \\
 T^{a}_{L,R ij} &=&   -\frac{i}{2} \left[\frac{1}{2}\epsilon^{abc}(\delta^{bi} \delta^{cj} -\delta^{bj} \delta^{ci} ) \pm (\delta^{ai} \delta^{4j} -\delta^{aj} \delta^{4i} )  \right], \nonumber \\
 T^{\alpha}_{ij} &=&-\frac{i}{\sqrt{2}} ( \delta^{\alpha i} \delta^{5j} -\delta^{\alpha j} \delta^{5i}),    
 \eea 
 where $\hat{a} =1, ... , 5$, $a_{L,R} =1, 2, 3$, $\alpha =1, ..., 4$ and especially  $T^\eta \equiv T^{ \hat{5} }$  is the generator of $SO(2)_\eta$.
 

\section{$SU(4)$ generators} 
\label{App:SU4_Sp4_generator} 
In this appendix, we list the generators of $SU(4)$. The broken and unbroken generators satisfy the following identity:
 \bea
T^\alpha = -V T^{\alpha T} V^T  \quad  T^{\hat{\alpha}}  =  V T^{\hat{\alpha} T} V^T,
\eea
where $V$ is global symmetry breaking VEV in Eq.(\ref{eq:vacuum}) .
So the ten unbroken generators can be identified as~\cite{Cacciapaglia:2014uja} 
\bea
T^a _L =\frac{1}{2} \left( \begin{array}{cc}
 \sigma^a & 0\\
0& 0 \\ 
\end{array}  \right), \quad T^a _R =\frac{1}{2} \left( \begin{array}{cc}
 0 & 0\\
0&- \sigma^{a T} \\ 
\end{array}  \right),     
\eea 
which form a custodial symmetry $SU(2)_L \times SU(2)_R$ of $Sp(4)$, and the remaining four generators are
\bea
 \frac{1}{2\sqrt{2}} \left( \begin{array}{cc}
 0 & i \sigma^a \\
-i \sigma^a & 0 \\ 
\end{array}  \right)   \quad  \frac{1}{2\sqrt{2}} \left( \begin{array}{cc}
 0 &   \mathds{1}_2 \\
 \mathds{1}_2 & 0 \\ 
\end{array}  \right).  
\eea
The five broken generators are 
\bea
T^{\hat{a}} &=&  \frac{1}{2\sqrt{2}} \left( \begin{array}{cc}
 0 &  \sigma^a \\
 \sigma^a & 0 \\ 
\end{array}  \right), \;  T^{\hat{4}} =\frac{1}{2\sqrt{2}} \left( \begin{array}{cc}
 0 &  i \mathds{1}_2  \\
 -i \mathds{1}_2 & 0 \\  
\end{array}  \right),  \nonumber \\
 \mbox{and}  \; T^{\eta } &=&  \frac{1}{2\sqrt{2}} \left( \begin{array}{cc}
 \mathds{1}_2 &  0 \\
0 & -\mathds{1}_2 \\  
\end{array}  \right),
\eea 
where $T^\eta$ associated with the singlet $\eta$ is the generator of a $U(1)$ symmetry which we denote as $U(1)_\eta$.

\section{The map from $SU(4)$ to $SO(6)$}
\label{App:Map}
The $SU(4)$ is automorphism with $SO(6)$ so in this section we present the correspondence between the representation of this two groups. It is well known that the subgroup $SO(4) \times SO(2)_\eta$ of $SO(6)$ is automorphism to the subgrounp $SU(2)_L \times SU(2)_R \times U(1)_\eta$ of $SU(4)$. The two spinor representations of $SO(4)$ correspond to the $SU(2)_L$ and $SU(2)_R$ fundamental representation respectively and the two ones of $SO(2)_\eta$ correspond to $U(1)_\eta$ self-representations which are conjugate with each other. So the spinor representation $\bf 4_{spin}$ of $SO(6)$ can be decomposed into the fundamental representations of $SU(2)_L$  and $SU(2)_R$ which take the opposite charge of $U(1)_\eta$ denoted as $x$:
\bea
4_{spin} =(2,1)_x +(1,2)_{-x}.
\eea 
So the spinor representation corresponds to the fundamental representation of $SU(4)$. The fundamental representation $\bf 6_F$ of $SO(6)$ can be decomposed under subgroup $SO(4) \times SO(2)_\eta$ as 
\bea
6_F =(4,1) +(1,2). 
\eea 
Since the fundamental representation of $SO(2)_\eta$ can be decomposed under $U(1)_\eta$ as $\bf 2 =1_x +1_{-x}$, the fundamental representation $\bf 6_F$ is decomposed under $SU(2)_L \times SU(2)_R \times U(1)_\eta$ as:
\bea
6_F =(2,2)_0 + (1,1)_x +(1,1)_{-x},
\eea
which exactly corresponds to the two indexes antisymmetric representation $\bf 6$ of $SU(4)$.

\section{Realization of $Z_2$ in Gauge Sector}
\label{App:Z_2_gauge}
For the gauge sector we can also apply above theory to realize the neutral naturalness. We gauge two subgroup of $H_1$ and $H_2$ with gauge bosons $W_\mu \equiv W^a_\mu T^a $ and $\tilde{W}_\mu \equiv \tilde{W}^{\tilde{a}}T^{\tilde{a}}$. So the gauge interaction for pNGBs is 
\bea
\mathcal{L} \supset  g\Sigma^T W_\mu W_\mu \Sigma +\tilde{g} \Sigma^T \tilde{W}_\mu \tilde{W}_\mu \Sigma.  
\eea 
So if $g =\tilde{g}$, the Lagrangian is invariant under $Z_2$ transformation $P^W$ 
\bea
W_\mu ^a T^a \to P^W \tilde{W}_\mu ^{\tilde{a}} T^{\tilde{a}}  P^{W \dagger}, ~ \tilde{W}_\mu ^{\tilde{a}} T^{\tilde{a}} \to P^W W_\mu ^a T^a  P^{W \dagger}.
\eea 
If the $P^W$ is the product of two $Z_2$ symmetry $P^W =P_0 ^W P_1$ with $[P_0 ^W, P_1 ]=0$, the Lagrangian is invariant under the following exchanging symmetry 
\bea
W_\mu ^a \leftrightarrow \tilde{W}_\mu ^{\tilde{a}} \quad \sin \left( \frac{\pi^i}{f} \right) \leftrightarrow \cos \left( \frac{\pi^i}{f} \right).  
\eea

\section{The Gauge Sector}
\label{App:Gauge} 
Like QCD, the composite Higgs potential from gauge interaction should be convergent at high energy scale because of its composite nature. In order to realize its high energy behavior the vector resonances should be introduced and impose Weinberg sum rules~\cite{Weinberg:1967kj} on these resonances spectrum and decay constants.  we introduce a set of vector resonances through hidden local symmetry~\cite{Bando:1987br}, where the vector resonances $\rho_\mu$, in the adjoint representation  of $SO(5)$, transform non-linearly, while the axial resonances $a_\mu$, in the fundamental representation  of $SO(5)$, transform homogeneously.  Under a global $SO(6)$ transformation $ {\bf g}$, we have 
\bea \label{eq:vector_resonaces}
\rho_\mu &=&  \rho^a _\mu T^a \quad  \rho_\mu \to  {\bf h} \rho_\mu {\bf h} ^\dagger  + \frac{i}{g_\rho} {\bf h } \partial_\mu {\bf h}^\dagger \nonumber \\ 
a_\mu &=& a^{\hat{a}} _\mu T^{\hat{a}}  \quad   a_\mu \to  {\bf h} a_\mu {\bf  h }^\dagger,   
\eea
where $T^{\hat{a}}$($T^a$)  is (un-)broken generators.  
So at leading order in derivatives, the general Lagrangian allowed by Eq.~(\ref{eq:vector_resonaces}) is~\cite{Marzocca:2012zn} 
\bea
\mathcal{L}_\rho &=& -\frac{1}{4} \mbox{Tr}[\rho_{\mu \nu} \rho_{\mu \nu} ] +\frac{f_\rho ^2}{2} \mbox{Tr}[(g_\rho \rho_\mu -E_\mu ^a T^a ) ^2]  \nonumber \\
\mathcal{L}_a &=&  -\frac{1}{4} \mbox{Tr}[a_{\mu \nu} a_{\mu \nu} ] +\frac{f_a ^2}{2\Delta^2} \mbox{Tr}[(g_a a_\mu -\Delta d_\mu ^{\hat{a}} T^{\hat{a}} )^2 ],   
\eea
where $i U^\dagger D_\mu U =d_\mu ^{\hat{a}} T^{\hat{a}} +E_\mu ^a T^a$,  $\rho_{\mu \nu}  =\partial_\mu \rho_\nu -\partial_\nu \rho_\mu -i g_\rho [\rho_\mu, \rho_\nu]$, $a_{\mu \nu}=\bigtriangledown_\mu a_\nu -\bigtriangledown_\nu a_\mu$ and $\bigtriangledown_\mu =\partial_\mu -i E_\mu ^a T^a$.
After integrating out the heavy resonances at tree level, the $SO(6)$ invariant Lagrangian, at quadratic order in the gauge fields and in momentum space, is  
\bea
\mathcal{L} ^{eff} &=& \frac{P_t^{\mu \nu}}{2}  \left( (\Pi_0(p^2) ) \mbox{Tr}[A_\mu A_{\nu}]    +\Pi_1(p^2) \Sigma^T A_\mu A_\nu \Sigma  \right. \nonumber \\ 
    &-& \left. p^2( W^a _\mu W^a _\mu +B_\mu B_\mu  + B_\mu ^\prime B_\mu ^\prime )    \right), 
\eea  
where $A_\mu =g W_\mu ^a T^a_L +g^\prime B_\mu T_R ^3 + g_1 B_\mu ^\prime T_\eta$, $P_t ^{\mu \nu} = g^{\mu \nu} - p_\mu p_\nu/p^2$ is the projector on transverse field configurations and $\Pi_{0,1}$ are form factors. From above Lagrangian, we get the most general effective Lagrangian for gauge bosons with explicit dependence on the Higgs field:
\bea
\mathcal{L}^{eff} &=& \frac{P_t^{\mu \nu}}{2}\left( g^2\Pi_0 ^W  W^a _\mu  W^a _\mu  + g^2\Pi_1 \frac{s_h ^2}{4}(W_\mu ^1 W_\mu^1 +W_\mu ^2 W_\mu ^2     ) \right. \nonumber \\     
    &+&\left.  g_1^2 (\Pi_0 ^\eta  + \Pi_1 \frac{c_h ^2}{2}  ) B_\mu ^\prime B_\mu ^\prime + g^{\prime 2} \Pi_0 ^B  B_\mu B_\mu \right. \nonumber \\ 
    &+& \left. \Pi_1 \frac{s_h ^2}{4} (g^\prime B_\mu - g W_\mu ^3 )(g^\prime B_\mu - g W_\mu ^3 )  \right), 
\eea
 where $\Pi_0 ^W = -\frac{p^2}{g^2} +  p^2 \frac{f_\rho ^2 }{p^2 -m_\rho ^2} $, $\Pi_0 ^{\eta} =\Pi_0 ^W(g \to g^\prime ) $, $\Pi_0 ^B  =\Pi_0 ^W(g \to g_1)$ and $\Pi_1 = f^2 + 2p^2 [{f_a ^2}/(p^2 -m_a ^2 )-{f_\rho ^2}/(p^2 -m_\rho ^2)] $. Here we define the mass parameters 
 \bea
 m_\rho ^2 = f_\rho ^2 g_\rho ^2 \quad m_a ^2 = \frac{f_a ^2 g_a ^2 }{\Delta ^2}.
 \eea 
 So it is easy to get the Higgs potential at one loop level, integrating out the gauge fields and going to Euclidean momenta,
 \bea
 V_g(h) &=& \frac{3}{2} \int \frac{d^4 p_E}{(2\pi)^4} \left(2 \mbox{log} [ \Pi_0 ^W + \Pi_1 \frac{s_h ^2}{4} ] + \mbox{log}[ \Pi_0 ^\eta + \Pi_1 \frac{c_h ^2}{2} ] \right.  \nonumber \\
   &+& \left. \mbox{log}[\Pi_0 ^B \Pi_0 ^W +\Pi_1 \frac{s_h ^2}{4}(\Pi_0 ^B +\Pi_0 ^W ) ]   \right ).
 \eea  
 It is straightforward to get the form of the gauge contribution to Higgs potential at order of gauge couplings square
 \bea
 \label{eq:Vg2}
 V_g \simeq \frac{3}{2(4\pi)^2 } \int_0  dp_E ^2 p_E^2 \left[ \left(\frac{3 \Pi_1}{\Pi_0 ^W}  +\frac{\Pi_1}{\Pi_0 ^B} \right) \frac{ s_h ^2}{4} + \frac{\Pi_1}{\Pi_0 ^\eta} \frac{c_h ^2 }{2} \right].    
 \eea
  
The behavior of form factors at high energy scale depends on the underlying strong dynamics. For composite Higgs model based on fermionic UV completion, according to operator product expansion, the scalar operator with the lowest dimension  contributes to form factor $\Pi_1$ is $\mathcal{O}_6 =\psi^\dagger \Gamma_1 \psi \psi^\dagger \Gamma_2 \psi $ which is dimension $6$, where $\psi$ denote preon $\psi_i$ and $\Gamma_{1,2}$ are matrices in flavor, hyper-color and Lorentz index, so its high energy behavior can be estimated as~\cite{Contino:2010rs}: 
 \bea
 \label{eq:OPE}
 \Pi_1(p_E) = p_E^2( C(p_E) \langle \mathcal{O}_6 \rangle +... ) =p_E ^2 \left( \frac{\delta }{p_E ^6} + \mathcal{O} \left(\frac{1}{p_E ^8} \right)  \right).
 \eea
Since the form factors $\Pi_0 ^{W, B, \eta}$ correspond to gauge boson kinetic term, they grows as $p_E^2$ at large energy region. So according to Eq.\ref{eq:Vg2}, it is easy to see that the Higgs potential $V_g$ is convergent. The behavior at large $p_E$ in Eq.~(\ref{eq:OPE}), $\Pi_1(p_E) \sim {1}/{p_E^4 } +\mathcal{O}({1}/{p_E ^6})$, implies two sum rules on the spectrum of spin-1 resonances mass and decay constants: 
\bea
f_\rho ^2 -f_a ^a =\frac{ f^2}{2} \quad f_\rho ^2 m_\rho^2 -f_a ^a m_a ^2 =0, 
\eea
which are the famous Weinberg sum rules~\cite{Weinberg:1967kj}. We impose the above sum rules and setting $f_\rho =f$ for simplicity, the leading Higgs potential from electroweak gauge bosons and $B_\mu ^\prime$ loops is
\bea 
\label{eq:Vg}
V_g \simeq \frac{3f^2(3g^2+g^{\prime 2} -  2g_1 ^2 )m_\rho ^2 \ln 2}{64\pi^2} s_h ^2.
\eea

\end{document}